\documentclass[11pt]{article}
\pdfoutput=1

\usepackage{cancel,slashed}
\usepackage{bbm}
\usepackage{bm}
\usepackage{amsmath,amsfonts,amssymb,mathtools,nicefrac,nccmath,cases}
\usepackage{graphicx}
\usepackage{cite}
\usepackage{skak}

\usepackage{dsfont}
\usepackage{latexsym}
\usepackage{color}
\usepackage[hyperfootnotes=false,linktocpage]{hyperref}
\usepackage{transparent}
\usepackage[hang,flushmargin]{footmisc}
\usepackage{blkarray}
\usepackage{multirow}
\usepackage{empheq}
\usepackage{comment}
\usepackage{wasysym}
\usepackage{tikzsymbols}

\usepackage{caption}
\usepackage{subcaption}
\usepackage[normalem]{ulem}

\newcommand{\bmat}{\left(\begin{array}}
\newcommand{\emat}{\end{array}\right)}

\def\R{\mathbbm{R}}

\def\a {\alpha}
\def\b {\beta}

\def\Tr{\text{Tr}}

\def\1{{\bf 1}}
\def\2{{\bf 2}}
\def\3{{\bf 3}}
\def\4{{\bf 4}}
\def\6{{\bf 6}}

\def\targ#1#2{\genfrac{[}{]}{0pt}{}{#1}{#2}}
\def\targ2#1#2{\genfrac{}{}{0pt}{}{#1}{#2}}

\definecolor{mygr}{rgb}{0,0.6,0}
\definecolor{mygrey}{rgb}{0,0.1,0.2}
\definecolor{myblue}{rgb}{0,0.5,0.9}
\definecolor{myblue2}{rgb}{0,0.5,0.5}
\definecolor{myblue3}{rgb}{0,0.7,0.9}
\definecolor{myblue4}{rgb}{0,0.6,0.6}
\definecolor{myorange}{rgb}{1,0.5,0}
\definecolor{mypurple}{rgb}{0.6,0,1}
\definecolor{mygolden}{rgb}{1,0.8,0.2}
\definecolor{mycyan}{rgb}{0,1,1}
\definecolor{mymagenta}{rgb}{1,0,1}
\definecolor{mykiwi}{rgb}{0.8,1,0.5}
\definecolor{mybrown}{cmyk}{0.14, 0.42, 0.56, 0.2}
\definecolor{myturq}{cmyk}{0.99, 0, 0.2, 0.4}
\definecolor{myaubergine2}{cmyk}{0.4, 0.5, 0, 0.1}
\definecolor{myaubergine}{cmyk}{0.6,0.85,0,0}
\definecolor{CycleGreen}{cmyk}{0.52,0,1,0}
\definecolor{CycleBrown}{cmyk}{0, 0.4, 0.9, 0.2}

\DeclareFontFamily{U}{rcjhbltx}{}
\DeclareFontShape{U}{rcjhbltx}{m}{n}{<->rcjhbltx}{}
\DeclareSymbolFont{hebrewletters}{U}{rcjhbltx}{m}{n}

\DeclareMathSymbol{\lamed}{\mathord}{hebrewletters}{108}
\DeclareMathSymbol{\mem}{\mathord}{hebrewletters}{109}
\DeclareMathSymbol{\ayin}{\mathord}{hebrewletters}{96}
\DeclareMathSymbol{\tsadi}{\mathord}{hebrewletters}{118}
\DeclareMathSymbol{\qof}{\mathord}{hebrewletters}{113}
\DeclareMathSymbol{\resh}{\mathord}{hebrewletters}{114}
\DeclareMathSymbol{\pe}{\mathord}{hebrewletters}{112}
\DeclareMathSymbol{\pesofit}{\mathord}{hebrewletters}{80}
\DeclareMathSymbol{\samekh}{\mathord}{hebrewletters}{115}
\DeclareMathSymbol{\tav}{\mathord}{hebrewletters}{116}
\DeclareMathSymbol{\vav}{\mathord}{hebrewletters}{119}
\DeclareMathSymbol{\het}{\mathord}{hebrewletters}{120}
\DeclareMathSymbol{\yod}{\mathord}{hebrewletters}{121}
\DeclareMathSymbol{\zayin}{\mathord}{hebrewletters}{122}
\DeclareMathSymbol{\alephdot}{\mathord}{hebrewletters}{128}
\DeclareMathSymbol{\tsadisofit}{\mathord}{hebrewletters}{90}
\DeclareMathSymbol{\shin}{\mathord}{hebrewletters}{152}

\newcommand{\LM}[1]{{\bf[{\color{blue}LM:} #1]}}

\def\CN {{\cal N}}

\def\CK {{\cal K}}

\def\be{\begin{equation}}
\def\ee{\end{equation}}
\def\bea{\begin{eqnarray}}
\def\eea{\end{eqnarray}}
\def\bes{\begin{subequations}}
\def\ees{\end{subequations}}
\def\raw{\rightarrow}

\def\oh{\frac{1}{2}}
\def\CG {{\cal G}}

\usepackage{multicol}
\usepackage{float}
\usepackage{caption}
\def\p {{\partial}}

\def\g {{\gamma}}

\def\vee {{\varepsilon}}

\newcommand{\cK}{\mathcal{K}}

\newcommand{\cN}{\mathcal{N}}

\newcommand{\cR}{\mathcal{R}}
\newcommand{\cT}{\mathcal{T}}

\newcommand{\IR}{\mathbb{R}}

\newenvironment{eqn*}{\begin{equation*}\begin{aligned}}{\end{aligned}\end{equation*}\noindent}

\makeatletter
\newsavebox\myboxA
\newsavebox\myboxB
\newlength\mylenA

\newcommand*\xoverline[2][0.75]{%
\sbox{\myboxA}{$\m@th#2$}%
\setbox\myboxB\null
\ht\myboxB=\ht\myboxA%
\dp\myboxB=\dp\myboxA%
\wd\myboxB=#1\wd\myboxA
\sbox\myboxB{$\m@th\overline{\copy\myboxB}$}
\setlength\mylenA{\the\wd\myboxA}
\addtolength\mylenA{-\the\wd\myboxB}%
\ifdim\wd\myboxB<\wd\myboxA%
   \rlap{\hskip 0.5\mylenA\usebox\myboxB}{\usebox\myboxA}%
\else
    \hskip -0.5\mylenA\rlap{\usebox\myboxA}{\hskip 0.5\mylenA\usebox\myboxB}%
\fi}
\makeatother

\begin{document}
\pagestyle{plain}

\makeatletter
\@addtoreset{equation}{section}
\makeatother
\renewcommand{\theequation}{\thesection.\arabic{equation}}

\pagestyle{empty}
\rightline{IFT-UAM/CSIC-22-81}
\vspace{0.5cm}
\begin{center}
\Huge{{EFT strings and emergence} 
\\[10mm]}
\normalsize{Fernando Marchesano\,$^{1}$  and Luca Melotti}\,$^{1,2}$\\[12mm]
\small{
${}^{1}$ Instituto de F\'{\i}sica Te\'orica UAM-CSIC, c/ Nicol\'as Cabrera 13-15, 28049 Madrid, Spain \\[2mm] 
${}^{2}$ Departamento de F\'{\i}sica Te\'orica, Universidad Aut\'onoma de Madrid, 28049 Madrid, Spain
\\[10mm]} 
\small{\bf Abstract} \\[5mm]
\end{center}
\begin{center}
\begin{minipage}[h]{15.0cm} 

We revisit the Emergence Proposal in 4d $\CN=2$ vector multiplet sectors that arise from  type II string Calabi--Yau compactifications, with emphasis on the role of axionic fundamental strings, or EFT strings. We focus on large-volume type IIA compactifications, where EFT strings arise from NS5-branes wrapping internal four-cycles, and consider a set of infinite-distance moduli-space limits that can be classified in terms of a scaling weight $w=1,2,3$. It has been shown before how one-loop threshold effects of an infinite tower of BPS particles made up of D2/D0-branes generate the asymptotic behaviour of  the gauge kinetic functions along limits with $w=3$. We extend this result to $w=2$ limits, by taking into account D2-brane multi-wrapping numbers. In $w=1$ limits the leading tower involves EFT string oscillations, and one can reproduce the behaviour of both weakly and strongly-coupled $U(1)$'s independently on whether the EFT string is critical or not, by assuming that charged modes dominate the light spectrum.

\end{minipage}
\end{center}
\newpage
\setcounter{page}{1}
\pagestyle{plain}
\renewcommand{\thefootnote}{\arabic{footnote}}
\setcounter{footnote}{0}


\tableofcontents


\section{Introduction}
\label{s:intro}

The Swampland Distance Conjecture (SDC) \cite{Ooguri:2006in} is one of the key proposals driving the current research on the general constraints that Quantum Gravity imposes on Effective Field Theories (EFTs), also known as Swampland Programme \cite{Vafa:2005ui} (see \cite{Brennan:2017rbf,Palti:2019pca,vanBeest:2021lhn,Grana:2021zvf,Harlow:2022gzl} for reviews). This is not only because of its central position in the growing web of conjectures that threads the programme, but also due to its ability to constrain high energy physics models involving large field excursions. 

Our knowledge of the physical principle underlying the conjecture has deepened as it has been tested in more involved settings. A crucial step in this quest has been the analysis of 4d $\CN=2$ theories obtained from compactifying type II string theory on a Calabi--Yau (CY) manifold. This class provides a plethora of examples, most of them leading to a moduli space with an intricate structure of infinite distance limits. Particularly enlightening is the vector multiplet sector of type IIB CY compactifications, of which we know the moduli space metric obtained after integrating out all massive states. This case was analysed in \cite{Grimm:2018ohb}, where it was shown that at each infinite distance limit there is an infinite tower of states that become massless at an exponential rate, as predicted by the SDC. In the case of \cite{Grimm:2018ohb} the tower of states is made up of charged BPS particles obtained from wrapping D3-branes on  three-cycles of the Calabi--Yau, and the analogy with \cite{Strominger:1995cz} served to sustain the so-called Emergence Proposal \cite{Harlow:2015lma,Grimm:2018ohb,Heidenreich:2018kpg}. 

In short, the Emergence Proposal turns around the logic of the SDC: it is because an infinite tower of particles becomes massless exponentially fast that one observes an infinite distance singularity in moduli space. This expectation can be substantiated by computing the one-loop corrections to the field space metric induced by the tower, including only those states that lie below the species scale, and checking whether the asymptotic behaviour of the correction and that of the exact metric match. Since in \cite{Grimm:2018ohb} the  tower elements are charged and satisfy the Weak Gravity Conjecture \cite{Arkani-Hamed:2006emk}, the same effect serves to explain the vanishing asymptotic behaviour of $U(1)$ couplings along these  trajectories \cite{Heidenreich:2017sim}. In 4d $\CN=2$ settings such a relation between  the metric and the gauge kinetic function follows from general supersymmetry arguments, so in order to test the Emergence Proposal one may focus on the asymptotic behaviour of one or the other. It was found in \cite{Grimm:2018ohb} that the Emergence Proposal explains the asymptotics of the gauge kinetic function diagonal components for a large class of limits, while a mismatch was found in some of them. The analysis was carried out by first classifying the limits in terms of the singularity index $n$, with a partial mismatch found for the case $n=2$. 

The aim of this paper is to revisit the Emergence Proposal with the current knowledge of type II CY compactifications. We focus on the vector multiplet sector of type IIA compactifications at large volume. This regime displays a simple moduli space metric, while still captures to great extent the different classes of infinite distance limits \cite{Grimm:2018cpv,Corvilain:2018lgw}. It also allows one to have a better control over the set of light states that are present in each limit, which is crucial to test the Emergence Proposal. In fact, it is in the type IIA setup where the authors of \cite{Lee:2019wij} showed that the light states of some $n=1$ limits can be packaged in the spectrum of an asymptotically tensionless dual critical string, which led them to formulate the Emergent String Conjecture. 

It was then realised that asymptotically tensionless strings could provide an organising principle to classify 4d infinite distance limit endpoints. Indeed, the physics of fundamantal axionic strings, dubbed EFT strings in \cite{Lanza:2020qmt,Lanza:2021udy,Lanza:2022zyg}, captures several key properties of those infinite distance trajectories that are described when approaching their core in their backreacted solution. Based on this, it was proposed a simple relation between the mass scales set by the lightest tower of states $m_*$ and by the EFT string tension $\sqrt{\cal T}$, captured by an integer $w$ dubbed scaling weight. In the case of type II CY compactifications $w$ corresponds to the singularity index $n$, and $w=1$ limits are expected to correspond to emergent string limits in the sense of \cite{Lee:2019wij}, in which the modes of a critical string dominate the light spectrum. 

Our approach will be to use all these results to perform a more detailed analysis of the Emergence Proposal in type IIA CY compactifications. We  focus our attention on infinite distance limits defined by EFT strings, dubbed EFT string limits. We  first revisit $w=n=2$ limits and show that, upon a careful analysis of the light spectrum of BPS particles, one obtains a corrected asymptotic behaviour for the species scale and a modified threshold correction to the gauge kinetic functions that allows to reproduce the asymptotic behaviour of all its axion-independent components. From here we also draw an important lesson: in our setup the leading and largest tower of light states must be populated by charged particles in order to reproduce the asymptotic behaviour of the gauge couplings via the Emergence Proposal. This result could be expected for the vector multiplet sector of $\CN=2$ supergravity theories, since there the metric and the gauge kinetic functions are simply related, and so one needs charged particles in order to simultaneously generate their divergences via one-loop corrections. 

We then turn our attention to $w=1$ limits which, as mentioned, are expected to be emergent string limits. In that case the computation of the species scale must be dominated by the EFT string oscillation modes and, if these modes are charged, the same applies to the gauge kinetic function corrections. Via the anomaly inflow argument used in \cite{Heidenreich:2021yda} we show that the EFT string modes are indeed charged, and then compute the one-loop corrections using a general Ansatz for the degeneracy of string modes, assuming that charged modes dominate the string spectrum. We then find that one reproduces the expected asymptotic behaviour of the gauge kinetic function, matching not only weakly-coupled $U(1)$'s, but also those that are driven towards strong coupling along the limit. Remarkably, the same result is recovered if one assumes that the EFT string is non-critical, and so the BPS D-brane particles dominate the light spectrum. It thus seems that, if our assumptions are correct, there is no preferred option for the Emergence Proposal.

The rest of the paper is organised as follows. Section \ref{s:scaling} describes the type IIA EFT string limits that we analyse and their classification in terms of the scaling weight $w$. Section \ref{s:simple} analyses a simple example that illustrates the differences between $w=3$ and $w=2$ limits in terms of particle towers.  It then shows how the Emergence Proposal reproduces the asymptotic gauge kinetic functions, a computation that is extended to general $w=2$ limits in section \ref{s:w2}. Section \ref{s:eslim} addresses emergence for $w=1$ limits, both in scenarios where the EFT string is critical and non-critical. Section \ref{s:examples} contains further examples that illustrate our general results and section \ref{s:conclu} our conclusions. Finally, appendix \ref{ap:democratic} describes the dimensional reduction of the 10d democratic type IIA supergravity action on a Calabi--Yau manifold, and appendix \ref{ap:K3} the gauge boson kinetic terms including all curvature corrections. 

\bigskip

{\em Note added:} At the latest stages of this project we became aware of \cite{Castellano:2022bvr}, which explores similar ideas in different contexts. We would like to thank the authors for sharing with us their results, as they have served to sharpen ours. 


\section{EFT strings and infinite distance limits}
\label{s:scaling}

Let us consider type IIA string theory compactified on a Calabi--Yau three-fold $X_6$ 
\be
ds^2 = ds_{\R^{1,3}}^2(x) + ds_{X_6}^2(y)\, ,
\ee
whose metric is characterised by the periods of its holomorphic three-form $\Omega$ and its K\"ahler form $J$. The latter can be expanded as
\be
J = t^a(x) \omega_a\, ,
\ee
where $\ell_s^{-2} \omega_a$ is a basis of integral harmonic two-forms Poincar\'e dual to a basis of Nef divisors, with $\ell_s = 2\pi \sqrt{\a'}$ the string length. We may use the same basis to dimensionally reduce the B-field and RR potentials
\be
B = b^a(x) \omega_a + \dots\, , \qquad C_1 = A_1^0(x)\, , \qquad C_3 = \hat{A}_1^a(x) \wedge \omega_a + \dots\, ,
\label{BCs}
\ee
where the dots represent terms of the expansion that are irrelevant for the vector multiplet sector. Here $T^a =  b^a + i t^a$ and $A_1^0$, $\hat{A}_1^a$ represent 4d complex scalars and vector bosons, respectively. Upon dimensional reduction of the standard type IIA 10d supergravity action one recovers the following piece of 4d $\CN=2$ Lagrangian describing the vector multiplet sector \cite{Ferrara:1988ff,Andrianopoli:1996cm}
\be
S_{\rm 4d}^{\rm VM} =  \frac{1}{2\kappa_{4}^2} \int_{\R^{1,3}} R * \mathbbm{1} - \frac{1}{2} G_{AB} dT^A \wedge * d\bar{T}^B + I_{AB} F^A \wedge *_4 F^B + R_{AB} F^A \wedge F^B ,
\label{SVM}
\ee
where $A = (0,a)$ and $F^A$ are integrally-quantised two-form field strengths associated to the U(1) gauge bosons in \eqref{BCs}. In particular, $F^0 = dA_1^0$ represents the field strength of the graviphoton, and $F^a = dA_1^a \equiv d(\hat{A}_1^a - b^a A_1^0)$. The kinetic terms and gauge couplings are specified by
\be
G_{AB} = \left(
\begin{array}{cc}
0 & \\ 
& 4 g_{ab}
\end{array}
\right) , \qquad 
g_{ab} = \frac{3}{2} \left( \frac{3}{2} \frac{\CK_a \CK_b}{\CK^2} - \frac{\CK_{ab}}{\CK} \right)  ,
\label{metric}
\ee
with $\CK_{abc} = \ell_s^{-6} \int \omega_a\wedge \omega_b \wedge \omega_c$ the triple intersection numbers of $X_6$, from where we build $\CK_{ab} = \CK_{abc} t^c$, $\CK_{a} = \CK_{abc} t^bt^c$ and $\CK =  \CK_{abc} t^at^bt^c = 6 {\cal V}_{X_6}$. We also have 
\be
I\, =\, - \frac{\CK}{6}
\left(
\begin{array}{cc}
1 +   4 g_{ab} b^ab^b & 4 g_{ab} b^b \\ 
4 g_{ab} b^b & 4 g_{ab}
\end{array}
\right) , \qquad
R\, =\, -
\left(
\begin{array}{cc}
 \frac{1}{3} \CK_{abc}b^ab^bb^c  & \oh  \CK_{abc}b^bb^c 
 \\  \oh  \CK_{abc}b^bb^c 
 & \CK_{abc}b^c 
\end{array}
\right) . 
\label{IandR}
\ee

In the large volume regime, single-field infinite distance limits have a simple description in terms of the K\"ahler cone, namely by considering the moduli space trajectories
\begin{equation}
t^a = e^a \phi, \quad \text{with} \quad \phi \to \infty ,
\label{limita}
\end{equation}
and $e^a \in \mathbbm{N}$. Such limits take the compactification manifold to infinite volume, and so in order to keep the field trajectory within the vector multiplet moduli space one has to rescale the 10d dilaton accordingly
\begin{equation}
g_s (\phi) \sim {\cal V}_{X_6} (\phi)^{1/2} \to \infty .
\label{codila}
\end{equation}
In the following we will focus our attention on this class of limits  which, following the terminology of \cite{Lanza:2020qmt,Lanza:2021udy,Lanza:2022zyg} (see also \cite{Buratti:2021fiv,Angius:2022aeq,Marchesano:2022avb,Cota:2022yjw,Martucci:2022krl,Wiesner:2022qys}), we dub {\it EFT string limits}, for reasons to be discussed shortly. 

In the type IIA context, infinite distance limits have been classified in \cite{Corvilain:2018lgw,Lee:2019wij} from different perspectives. In general one obtains three different main types which, when applied to EFT string limits, have the following description:

\begin{enumerate}
    \item ${\bf k} \equiv \CK_{abc}e^ae^be^c\neq 0$, dubbed type IV$_d$ or $n=3$ limits in \cite{Corvilain:2018lgw}.
    \item ${\bf k} = 0$, but ${\bf k}_a \equiv \CK_{abc} e^be^c \neq 0$ for some $a$, dubbed type III$_c$ or $n=2$ limits in \cite{Corvilain:2018lgw}, and J-Class A limits in  \cite{Lee:2019wij}.
    \item ${\bf k}_a  = 0$, $\forall a$, dubbed type II$_b$ or $n=1$  limits in  \cite{Corvilain:2018lgw}, and J-Class B limits in \cite{Lee:2019wij}.
\end{enumerate}
For each of these limits the compactification manifold diverges, and so there is a tower of Calabi--Yau Kaluza--Klein modes that becomes asymptotically massless along \eqref{limita}. The spectrum of light modes and their scaling can be estimated by using the fibration structure that $X_6$ needs to posses in order to realise certain limits, as shown in  \cite{Lee:2019wij}. In particular, J-Class A limits must correspond to an elliptic fibration over a K\"ahler surface, while J-Class B limits of the form \eqref{limita} correspond to either a $K3$ or a $T^4$ fibered over a $\mathbbm{P}^1$ base. Finally, the subindices in II$_b$, III$_c$, and IV$_d$ correspond to a positive integer that classify subclasses of limits. As we will see, one can give these indices a physical meaning, in terms of towers of BPS particles that, just like the CY KK tower, become asymptotically massless as we proceed along \eqref{limita}. These towers are made up from D$(2p$)-branes wrapping internal $2p$-cycles of $X_6$ and, from the viewpoint of  the IR physics of the compactification, they are in general  more relevant than the light CY KK modes. 

The existence and importance of D-branes that appear as infinite towers of  BPS particles in 4d was emphasised in \cite{Grimm:2018ohb} in the mirror dual framework of infinite distance limits in type IIB complex structure moduli space. In this more general context, the presence of light towers of CY KK modes is less obvious, and so part of the motivation was to show that there is always a tower of particles that becomes asymptotically massless at an exponential rate compared to the 4d Planck scale, as predicted by the Swampland Distance Conjecture \cite{Ooguri:2006in}. In type IIA large-volume limits like \eqref{limita} such light states always contain D0-brane towers and D2/D0-bound states, made up from D2-branes wrapping complex curves on $X_6$ and with a non-trivial worldvolume flux ${\cal F} = B + \frac{\ell_s^2}{2\pi} F$. The subset of D2-branes that becomes massless depends on the particular limit that one takes, as worked out in \cite{Grimm:2018cpv,Corvilain:2018lgw} and illustrated below in some examples. 

It turns out that these D-brane towers are more relevant for IR physics than the CY KK tower for three reasons. First, their mass decreases either faster or at the same rate than the CY KK tower, so they would be noticed first from the viewpoint of an IR observer. At large volume the D0-brane tower is always the lightest one, and sets the maximum cut-off  $m_*$ of the 4d EFT. Such a  scale separation is parametric from all the other towers for $n=3$ limits, see table \ref{tab:scaling_L(phi)}. It follows that such limits can be seen as decompactification limits to M-theory on $X_6$ \cite{Lee:2019wij}.

\vspace*{.5cm}
\setlength{\arrayrulewidth}{0.2mm}
\renewcommand{\arraystretch}{1.25}
\begin{table}[H]
\begin{center}
\begin{tabular}{|c|c|c|c|c|c|}
\hline
 $n=w$  & $m_{\rm KK}$ & $m_*=m_{\text{D0}}$ & $m_{\text{D2}}$ & $\Lambda_{\text{NS5}}=\sqrt{{\cal T}}$ & $\Lambda_{\text{sp}}$ \\
\hline
3 & $\phi^{-\frac{1}{2}}$  & $\phi^{-\frac{3}{2}}$ & $\phi^{-\frac{1}{2}}$ & $\phi^{-\frac{1}{2}}$   & $\phi^{-\frac{1}{2}}$ \\
\hline
2  & $\phi^{-\frac{1}{2}}$  & $\phi^{-1}$ & $\phi^{-1}$  & $\phi^{-\frac{1}{2}}$ & $\phi^{-\frac{1}{3}}$ \\
\hline
1  & $\phi^{-\frac{1}{2}}$ & $\phi^{-\frac{1}{2}}$  & $\phi^{-\frac{1}{2}}$ & $\phi^{-\frac{1}{2}}$ & $\phi^{-\frac{1}{6}}$\\
\hline
\end{tabular}
\end{center}
\caption{Mass scales along infinite distance limits. Data collected from \cite{Grimm:2018ohb,Lee:2019wij,Lanza:2021udy}.
\label{tab:scaling_L(phi)}}
\end{table}

Second, they are charged under the $U(1)$'s of the theory. This is particularly significant given the asymptotic behaviour that the gauge kinetic function components have along \eqref{limita}. Particularly simple is the graviphoton kinetic term $I_{00}$, which presents a rather universal behaviour displayed in table \ref{tab:limits}. The diagonal components $I_{aa}$, for some specific directions different from the growing K\"ahler modulus in \eqref{limita} are also displayed there. As these components diverge, the corresponding gauge couplings vanish, which would result into a global gauge symmetry. From this perspective, the appearance of an asymptotically massless tower of charged particles that triggers the 4d EFT  breakdown  can be interpreted as a protection mechanism against realising a global symmetry in a certain EFT field space limit 
\cite{Grimm:2018ohb}.
\setlength{\arrayrulewidth}{0.2mm}
\renewcommand{\arraystretch}{1}
\begin{table}[H]
\begin{center}
\begin{tabular}{|c|c|c|}
\hline
 $n=w$  & $I_{00}$ & $I_{aa}$   \\
\hline
3 & $\phi^3$ & $\phi$ \\
\hline
2  & $\phi^2$ & $\phi^2$  \\
\hline
1  & $\phi$ & $\phi$  \\
\hline
\end{tabular}
\end{center}
\caption{Leading behaviour of diagonal gauge kinetic function components along limits of the form \eqref{limita}. For $w=2$ the index $a$ is such that ${\bf k}_a  \neq 0$, and for $w=1$ any direction that can be written as ${\bf k}_{ab}\bar{e}^b$ for some $\bar{e}^a \in \mathbbm{N}$.
\label{tab:limits}}
\end{table}

Third, they transform under the shift symmetries of the theory. These are of the form
\begin{equation}
    b^a \to b^a + \lambda^a ,
    \label{shift}
\end{equation}
with $\lambda^a \in \mathbbm{R}$. This symmetry is manifest if we rewrite \eqref{SVM} as
\be
S_{\rm 4d}^{\rm VM} =  \frac{1}{2\kappa_{4}^2} \int_{\R^{1,3}} R * \mathbbm{1} - \frac{1}{2} G_{AB} dT^A \wedge * d\bar{T}^B - \tilde{I}_{AB} \tilde{F}^A \wedge *_4 \tilde{F}^B - \tilde{R}_{AB} \tilde{F}^A \wedge \tilde{F}^B ,
\label{SVMg}
\ee
where $\tilde{F}^0 = F^0$ and $\tilde{F}^a = F^a + b^a F^0$. In this new basis we have
\be
\tilde{I}\, =\,  \frac{\CK}{6}
\left(
\begin{array}{cc}
1  & 0 \\ 
0 & 4 g_{ab}
\end{array}
\right) , \quad
\tilde{R}\, =\, 
\left(
\begin{array}{cc}
 \frac{1}{3} \CK_{abc}b^ab^bb^c   & -\oh  \CK_{abc}b^bb^c  \\ 
- \oh  \CK_{abc}b^bb^c  & \CK_{abc}b^c
\end{array}
\right) ,
\label{IandRg}
\ee
and the $B$-field axions only appear as theta angles. Curvature and worldsheet instanton corrections will modify these expressions, with the latter inducing a dependence on $e^{i n_a T^a}$ and $e^{-i n_a \bar{T}^a}$, with $n_a \in \mathbbm{N}$. It is due to these terms the shift symmetry reduces to a gauge discrete symmetry \eqref{shift} in which $\lambda^a \in \mathbbm{Z}$, as expected. However, in the limit \eqref{limita} some instanton corrections are suppressed and the continuous shift symmetry $b^a \to b^a + c e^a$, $c\in \mathbbm{R}$, is recovered. Again, the presence of the asymptotically massless tower of states prevents realising such a global symmetry. 

It is this third feature that was used in \cite{Lanza:2021udy} to provide a physical realisation of the limits \eqref{limita} from a 4d EFT perspective, as it seems to be the one that is also present in settings with less supersymmetry. The key observation is that there is a physical EFT object, namely a 4d fundamental string with magnetic charges $e^a$ under axionic shifts, that realises the field space trajectory \eqref{limita} when approaching its core along its backreacted solution. Imposing the Weak Gravity Conjecture on this string allows to deduce an exponential decay rate for its tension along the limit \eqref{limita}, providing a rationale for the Swampland Distance Conjecture in 4d. Such objects were dubbed EFT strings in  \cite{Lanza:2020qmt,Lanza:2021udy,Lanza:2022zyg}, where they were proposed to characterise infinite distance trajectories endpoints in 4d EFTs.

In the case at hand, EFT strings are made up of NS5-brane wrapping Nef divisors of $X_6$, and they are in one-to-one correspondence with limits of the form \eqref{limita}. It is easy to check that a string wrapped on the divisor $e^aD_a$ dual to $\ell_s^{-2}e^a[\omega_a]$ has a tension ${\cal T}$ that vanishes like $\phi^{-1}$ along \eqref{limita}. If their oscillation modes form an infinite tower, they will compete with the other light towers that arise along such a limit. From table \ref{tab:scaling_L(phi)} one can see that the scale $\sqrt{{\cal T}}$ associated to a string tower is in general not the one with the fastest asymptotic rate. Nevertheless, it displays a simple relation with the behaviour of the scale of the leading tower  $m_* = m_{\rm D0}$:
\begin{equation}
    \frac{m_*^2}{M_{\rm P}^2} \sim \left(\frac{\cal T}{M_{\rm P}^2} \right)^w ,
    \label{scaling}
\end{equation}
where $w$ is an integer dubbed  scaling weight. It was proposed in \cite{Lanza:2021udy} that the asymptotic relation \eqref{scaling} is a general feature of 4d compactifications. In the case at hand, one can verify this behaviour with $w = n$ \cite{Lanza:2021udy}. In other words, the scaling weight corresponds to the singularity type of \cite{Grimm:2018ohb,Grimm:2018cpv,Corvilain:2018lgw}, that classifies infinite-distance limits. Table \ref{tab:scaling_L(phi)} also reveals that the EFT string scale is below the species scale \cite{Dimopoulos:2005ac,Dvali:2007hz,Dvali:2007wp}, computed in terms of the tower of D0's:
\be
\Lambda_{\rm sp} = \frac{M_{\rm P}}{\sqrt{S_{\rm tot}}}, \quad S_{\rm tot} = S_{\rm D0} = \frac{\Lambda_{\rm sp}}{\Delta m} \quad \Rightarrow \quad
\begin{cases}
\Lambda_{\rm sp} \sim (\Delta m)^{\frac{1}{3}} \sim \phi^{-\frac{n}{6}}\\
S_{\rm D0} \sim (\Delta m)^{-\frac{2}{3}} \sim \phi^{\frac{n}{3}}
\end{cases} .
\label{spD0}
\ee
As we will see, this estimate of the species scale must be corrected in some instances, but $\Lambda_{\rm NS5}$ will remain at or below $\Lambda_ {\rm sp}$, so it still makes sense as a fundamental EFT object. 

Either from \eqref{scaling} or table \ref{tab:scaling_L(phi)}, it follows that along limits with scaling weight $w=1$ the presence of EFT strings should be particularly relevant for IR physics. As emphasised in \cite{Lee:2019wij}, if the string tower asymptotes as $m_*$ and it displays the spectrum of a critical string, then their oscillation modes will dominate over any number of KK-like particle towers,  and one should understand the said trajectory as an equi-dimensional {\it emergent string limit}. It was indeed shown in \cite{Lee:2019wij} that the oscillation modes of an EFT string in $w=1$ limits with K3 fibre together with the asymptotically massless D-brane particles correspond to the spectrum of a critical string in a dual heterotic frame. The general expectation is that all EFT string limits with $w=1$ correspond to emergent string limits, although no general proof has been provided so far.

\subsubsection*{The scaling weight and monodromies}

One of the main ideas put forward in \cite{Lanza:2021udy} is that the physical properties of an EFT string characterise the asymptotic behaviour of the theory in the corresponding limit, and vice versa. For instance, the singularity type/scaling weight describes the asymptotic behaviour of the K\"ahler potential $K \sim - n \log \phi$ along \eqref{limita}. Geometrically, $n=w$ corresponds to the order of nilpotency of the log-monodromy transformation associated to such a limit. The  geometric picture involving monodromy actions was exploited in  \cite{Grimm:2018ohb,Grimm:2018cpv,Corvilain:2018lgw}, to find towers of charged BPS particles that become asymptotically massless along infinite distance limits. From the 4d viewpoint, these monodromies are the B-field axion shift $b^a \to b^a + n^a$ that one implements when circling around an NS5-brane wrapped on the divisor $n^aD_a$. In order for this shift to describe a discrete gauge symmetry of the theory, it must be accompanied by a relabelling of the particles charged under the $U(1)$ gauge symmetries, made up from D($2p$)-branes wrapping $2p$-cycles of $X_6$ and threaded by worldvolume fluxes. In particular, it implies a relabelling of the graviphoton charge of the D2/D0 bound states, that are the prototypical example of light BPS towers in large volume regimes. Microscopically, this relabelling corresponds to a simultaneous discrete shift in the vev of the B-field and the quantised worldvolume flux $F$ that leaves the D-brane field strength ${\cal F} = B + \frac{\ell_s^2}{2\pi} F$ invariant. Macroscopically, acting with the monodromy generates an infinite orbit of particles with fixed charge under $F^a$ and varying graviphoton charge. Using that these objects are BPS and the asymptotic behaviour of $m_{\rm D0}$ one can then show that if one element of the orbit is asymptotically massless, all of them are.

Monodromies act on charged particles because they act on the $U(1)$ field strengths $F^A$ themselves, as one can already see from \eqref{IandR}. A first hint of the monodromy structure is how the scaling weight/singularity type appears in the matrix $R$.  Indeed, a limit of the form \eqref{limita} is associated to the axionic direction $b^a \propto e^a$,  and such an axionic direction couples with power $n=w$ to the graviphoton topological term $F_2^0 \wedge F_2^0$. This sort of relation was observed in \cite{Heidenreich:2021yda} in a simple example. Here we see that the result is general in 4d ${\cal N}=2$ CY settings. For $w=3$ limits one should also be able to interpret it in terms of the physics of monopole {\it supergravity} strings that arise from  compactifications of M-theory on $X_6$  \cite{Katz:2020ewz}. 

A simple procedure to describe the monodromy structure underlying the vector multiplet sector is to construct a  pseudo-action that reproduces the equations of motion and Bianchi identities of \eqref{SVM}, and where the monodromy structure is manifest. This can be achieved by direct dimensional reduction of the 10d type IIA democratic action of \cite{Bergshoeff:2001pv}, as performed in Appendix \ref{ap:democratic}. The result is a Lagrangian of the form
\be \label{psSVM}
-\frac{1}{8\kappa_4^2} \int_{M_4}  {\bf {F}}^{\, t} {\cal R}^{\, t} {\cal G} {\cal R} \ast_4{\bf {F}}\, ,
\ee
where ${\bf {F}}^{\, t} = (F^0, F^a, F_{a} ,-F_{0})$ is a vector that contains the field strengths in \eqref{SVM} and their duals $F_{A} = \delta S_{\rm 4d}/\delta F^A$. The moduli dependence is contained in the matrices
\be \label{GandR}
{\cal G}\, =\, \frac{\CK}{6}
\left(
\begin{array}{cccc}
 1  \\ 
  & 4 g_{ab} \\ 
  && \frac{9}{\CK^2} g^{ab} \\
  & & &   \frac{36}{\CK^2}
\end{array}
\right)\, , \qquad 
{\cal R} = 
\left(
\begin{array}{cccc}
 1 & 0 & 0 & 0 \\ 
  b^a & \delta^{a}_{b}  & 0 & 0\\ 
 \oh  \CK_{abc}b^bb^c &  \CK_{abc}b^c &   \delta^{a}_{b} & 0 \\
\frac{1}{6} \CK_{abc}b^ab^bb^c  & \oh  \CK_{abc}b^bb^c& b^a & 1
\end{array}
\right)\, ,
\ee
where ${\cal G}$ can be understood as a saxion-dependent metric and ${\cal R}$ as an axion-dependent monodromy matrix. This expression is also familiar from the kinetic terms of three-forms \cite{Herraez:2018vae}, the reason being the one-to-one correspondence between 4d particles and membranes made up from D-branes wrapping internal cycles of $X_6$. In fact, using the same reasoning one can identify the scalar potentials present in $\CN=2$ gauged supergravities \cite{DAuria:1990qxt,Louis:2002ny} and the bilinear expressions for RR flux potentials derived in \cite{Bielleman:2015ina,Carta:2016ynn,Herraez:2018vae}.

Both in the case of four-form and two-form field strengths, the kinetic terms are subject to curvature corrections. These result in the following expression for the monodromy matrix \cite{Palti:2008mg,Escobar:2018rna}
\begin{equation}
    \bar{\cal R} = Q e^{b^a\bar{P}_a} = e^{b^aP_a}Q, 
    \label{monoexp}
\end{equation}
with $\bar{P}_a = Q^{-1} P_a Q$ and 
\begin{equation}    \label{PaQ}
    P_a = \begin{pmatrix} 0 & 0 & 0 & 0\\
    \delta_a &0 & 0 & 0\\
    0 & \CK_{abc} &0 & 0\\
    0 & 0 & \delta_a &0
    \end{pmatrix}, \qquad
   Q = \begin{pmatrix}
     1 & 0 & 0 & 0\\
     0 & \delta_a^b & 0 & 0\\
     K_a^{(2)} & - K_{ab}^{(1)} &  \delta_a^b &0 \\
     0 & - K_a^{(2)} & 0 & 1
    \end{pmatrix},
\end{equation}    
where
\be
K^{(1)}_{ab} = \frac{1}{2} {\cal K}_{aab} \, , \qquad K_a^{(2)} = \frac{1}{24\ell_s^6} \int_{X_6} c_2(X_6) \wedge \omega_a\, .
\label{Kcurv}
\ee
These corrections are encoded in the following curvature-corrected prepotential
\begin{equation} \label{corr_prep}
    {\cal F} = -\frac{1}{6} \cK_{abc}T^aT^bT^c + \oh K_{ab}^{(1)}T^aT^b + K_{a}^{(2)}T^a + \frac{i}{2} K^{(3)}, 
\end{equation}
where $K^{(3)} = \frac{\zeta(3)}{8\pi^3} \chi(X_6)$ is a curvature correction that does not modify the monodromy matrix, and that will be neglected in the following (see appendix \ref{ap:K3} for the exact expressions). Putting everything together we have the axion-dependent transformations
\begin{equation}
    b^a \to b^a +1 , \qquad {\bf {F}} \to e^{-\bar{P}_a} {\bf {F}},
\end{equation}
which extend those pointed out in \cite{Heidenreich:2021yda} to the present setup. They imply that the axion-invariant field strengths are given by the entries of $\bar{\cal R} {\bf {F}}$, instead of simply ${\bf {F}}$. This motivates using the \eqref{SVMg} if one is interested in axion-invariant quantities. Notice that in the basis $\tilde{F}^A$ the charges of particles under $\tilde{F}^a$ are not integrally quantised, as they depend on the axion vevs. Finally, adding curvature corrections modifies \eqref{IandRg} to
\be
\tilde{I}\, =\,  \frac{\CK}{6}
\left(
\begin{array}{cc}
1  & 0 \\ 
0 & 4 g_{ab}
\end{array}
\right) , \quad
\tilde{R}\, =\, 
\left(
\begin{array}{cc}
 \frac{1}{3} \CK_{abc}b^ab^bb^c  -2K_a^{(2)}b^a -K_{ab}^{(1)}b^a b^b  & -\oh  \CK_{abc}b^bb^c +K_a^{(2)} +K_{ab}^{(1)}b^b  \\ 
- \oh  \CK_{abc}b^bb^c  +K_a^{(2)} +K_{ab}^{(1)}b^b & \CK_{abc}b^c -K_{ab}^{(1)}
\end{array}
\right) .
\label{IandRgcurv}
\ee

\subsubsection*{Emergence and EFT strings}

The appearance of an infinite tower of asymptotically massless charged states prompted the authors of \cite{Grimm:2018ohb} to put forward the Emergence Proposal (see also \cite{Harlow:2015lma,Heidenreich:2017sim,Heidenreich:2018kpg,Ooguri:2018wrx,Palti:2019pca}). The proposal implies that the asymptotic behaviour of the gauge coupling constants and the field space metric can be understood as an IR effect obtained from integrating out an infinite tower of charged states. Their computations indeed reproduced the diagonal components $I_{00}$ and $I_{aa}$ in the cases $w=3$ and $w=1$ displayed in table \ref{tab:limits}, while some mismatch was found for $I_{aa}$ in the case $w=2$. In the next section we will revisit this mismatch in a specific example with single-field limits of type $w=3$ and $w=2$, and address how to reproduce the gauge kinetic matrix for the latter, by considering the tower of asymptotically massless states in more detail. We will then extend our results to more general $w=2$ limits in section \ref{s:w2}.  
For simplicity we will work in the gauge-invariant basis that corresponds to \eqref{IandRg}, where off-diagonal terms $\tilde{I}_{a0}$ involving the graviphoton vanish. A more detailed analysis of the terms $I_{a0}$ from the viewpoint of the Emergence Proposal  will be discussed in \cite{Castellano:2022bvr}. 

Moreover, even if the results of \cite{Grimm:2018ohb} reproduce the components $I_{00}$ and $I_{aa}$ in the case $w=1$, recall that these are limits in which the EFT string oscillations have the same asymptotic behaviour as the lightest tower of D-brane particles. From the picture drawn in \cite{Lanza:2021udy} and the results of \cite{Lee:2019wij}, the expectation is that these are emergent string limits in the sense of the Emergent String Conjecture \cite{Lee:2019wij}, and so the light EFT string corresponds to a critical string in a dual frame. In this case the string modes will dominate the light spectrum, and should modify the computation of one-loop threshold corrections to the gauge kinetic functions. To evaluate their effect, it is particularly relevant to determine if some modes are charged under the $U(1)$ gauge symmetries of the EFT.  It was argued in \cite{Heidenreich:2021yda} that they are indeed charged under $U(1)$ gauge symmetries in generic 4d $\CN=1$ compactifications. In section \ref{s:eslim} we apply the anomaly inflow argument of \cite{Heidenreich:2021yda} to our $\CN=2$ setting to argue that EFT string modes are also charged in this case, contributing to the one-loop corrections to the gauge coupling constants. Even in this drastically different setup, we find that one can still reproduce the asymptotic behaviour of the gauge coupling constants in such limits, by assuming a spectrum for the dual critical string dominated by charged modes.


\section{Emergence in a simple example}
\label{s:simple}

Let us consider a two-modulus example based on the Calabi--Yau studied in \cite{Candelas:1994hw}, which can be seen as the weighted projective space $X_6 = \mathbb{P}^{(1,1,1,6,9)}[18]$, and as an elliptic fibration over $\mathbbm{P}^2$ with $b_2(X_6) = 2$ and triple intersection numbers
\be
\label{intex}
{\cal K}_{111} = 9, \qquad {\cal K}_{112} = 3, \qquad {\cal K}_{122} = 1, \qquad {\cal K}_{222} = 0.
\ee
The large-volume infinite distance limits of this three-fold were worked out in \cite{Grimm:2018cpv}, together with their towers of asymptotically massless charged states. We have two different single-field limits:
\begin{enumerate}
\item $t^1 \rightarrow \infty \quad n=w=3$ ,
\item $t^2 \rightarrow \infty \quad n=w=2$ ,
\end{enumerate}
and we are interested in the asymptotic behaviour of the gauge kinetic functions along them
\be \label{gaugekin}
\tilde{I}_{00} = \frac{\cal K}{6}, \qquad  \tilde{I}_{ab} = \frac{\cal K}{6} 4 g_{ab}\, .
\ee
In the limit $t^1 \rightarrow \infty$  we have that ${\cal K} \sim \phi^3$ and $g_{ab} \sim \phi^{-2}$, and therefore
\begin{equation}
    \tilde{I}_{00} \sim \phi^3, \qquad \tilde{I}_{ab} \sim \phi,
    \label{Ilim1}
\end{equation}
while for $t^2 \rightarrow \infty$ we instead have ${\cal K} \sim \phi^2$ and $g_{11} \sim \text{const}$,  $g_{12} \sim g_{22} \sim \phi^{-2}$, resulting in 
\begin{equation}
    \tilde{I}_{00} \sim \phi^2, \qquad  \tilde{I}_{11}  \sim \phi^2, \qquad \tilde{I}_{12}  \sim \tilde{I}_{22} \sim \text{const.}
\label{Ilim2}
\end{equation}

Next, we would like to reproduce these asymptotic values using the same strategy as in \cite{Grimm:2018ohb}, namely by integrating out an infinite tower of charged states and taking into account their threshold corrections to the different components of the gauge kinetic function.  The correction we obtain by integrating out a tower of hypermultiplets with masses $m_k$ and charges $q_{k,A}$ is
\be
I_{AB}^{\rm IR} = I_{AB}^{\rm UV} - \sum_{k=0}^S \frac{8}{3\pi^2} q_{k,A}q_{k,B} \log \frac{\Lambda_{\rm UV}}{m_k}\, ,
\label{1loopI}
\ee
where $S$ is the number of particles in the tower below the species scale $\Lambda _{\rm sp} = \Lambda_{\rm UV}$. In each limit of our example we have a tower of BPS states, given by D2-branes wrapped on holomorphic 2-cycles $\Sigma_2$, whose masses can be computed as
\be
m_{\textbf{q}} = e^{\frac{K}{2}} \left|\int_{\Sigma_2}B+\frac{\ell_s^2}{2\pi}F+\text{i}J\right | = \frac{|q_1T^1 + q_2T^2 -\text{i}q_0|}{2\sqrt{3(t^1)^3+3(t^1)^2 t^2+ t^1(t^2)^2}}\, ,
\label{mq}
\ee
where $q_1$ and $q_2$ are the wrapping numbers of $\Sigma_2$ along the two effective curves that generate the Mori cone, see e.g. \cite[section 4.2]{Lanza:2021udy} for details, and $q_0$ is the D0-brane charge induced by the quantised worldvolume flux $F$. In general, one can consider a vector of integer charges as in \cite[section 4.6]{Grimm:2018cpv}
\be
\textbf{q} = (D6,D4_1,D4_2,D2_1,D2_2,D0) = (0,0,0,q_1,q_2,q_0)\, ,
\ee
where in the rhs we have only included those charges that contain BPS particles becoming asymptotically massless along the above large-volume limits.\footnote{In the gauge-invariant basis $\tilde{F}^A$ in which we are working, one should actually consider the charge vector
\begin{equation*}
   {\bf \tilde{q}} = (0,0,0,q_1,q_2,q_0+q_1b^1+q_2b^2)\, ,
\end{equation*}
where the D0-brane charge is not quantised. This does not make a substantial difference when computing threshold corrections to $\tilde{I}_{00}$ and $\tilde{I}_{ab}$, because the axion dependence is subleading. Notice that we are also ignoring the $b^a$ dependence in \eqref{mq}, and according to \cite{Castellano:2022bvr} both effects should cancel. Therefore, in practice one may phrase our discussion in terms of {\bf q}, as we will do for simplicity.} 
Using $Sl(2)$-orbit techniques, the following towers of BPS particles were found in \cite[section 4.6]{Grimm:2018cpv}
\begin{enumerate}
\item $t^1 \rightarrow \infty$ \be \label{q_1_ex} \textbf{q} = (0,0,0,9,3,-9k)\, , \ee
\item $t^2 \rightarrow \infty$ \be \textbf{q} = (0,0,0,1,0,-k)\, , \ee
\end{enumerate}
where $k \in \mathbb{Z}$. These two towers represent D2-branes wrapped on specific effective curves and with increasing D0-brane charge induced by the quantised worldvolume flux $F$. When one inputs this spectrum into the correction term in \eqref{1loopI}, one recovers two types of sums already considered in  \cite[section 6]{Grimm:2018ohb}
\be
\sum_{k=-S_{\rm D0}}^{S_{\rm D0}} \log \frac{\Lambda_{\rm UV}}{m_k} \sim S_{\rm D0} \sim \phi^{\frac{n}{3}}\, , 
\qquad \sum_{k=-S_{\rm D0}}^{S_{\rm D0}} k^2 \log \frac{\Lambda_{\rm UV}}{m_k} \sim S_{\rm D0}^3 \sim \phi^n\, ,
\ee
depending on whether the charge of the tower with respect to the gauge field we consider stays constant or is increasing. In both limits of our example the charges under the graviphoton $F^0$ are increasing with $k$, while the charges under the other gauge fields remain constant, so we find
\begin{enumerate}
\item $t^1 \rightarrow \infty$ : \qquad $\tilde{I}_{00} \sim \phi^3$ , \qquad $\tilde{I}_{ab} \sim \phi$ , 
\item $t^2 \rightarrow \infty$ : \qquad $\tilde{I}_{00} \sim \phi^2$ , \qquad $\tilde{I}_{11}  \sim \phi^{2/3}$ , \qquad $\tilde{I}_{12}  \sim \tilde{I}_{22} \sim$ const .
\end{enumerate}
As one can see, these computations reproduce the asymptotic behaviour \eqref{Ilim1} and \eqref{Ilim2} only partially, with a disagreement in the non-constant component $\tilde{I}_{11}$ along the limit $t^2 \rightarrow \infty$. This mismatch was already observed in \cite{Grimm:2018ohb} and, as we will now argue, it can be fixed via a more careful inspection of the species scale. 

Indeed, an important feature of the limits $n=3$ like $t^1 \to \infty$ is that the species scale $\Lambda_{\rm sp}$ as computed in \eqref{spD0} has the same scaling $\phi^{-1/2}$ as the the D2-brane $m_{\rm D2}$ and the EFT string scale $\sqrt{\cal T}$, see table \ref{tab:scaling_L(phi)}. This means that below this scale there is only a finite number of EFT string states or multi-wrapped D2-brane states in the theory. In other words, the D0's is the single tower of states  that becomes infinite asymptotically. Thus, one can consider them as the only tower contributing to the species scale, as assumed in \eqref{spD0}. Of course, there is also the tower of D2/D0 bound states with unit D2-brane charge, which is the one modifying the gauge kinetic function. However, this can be seen as a mass-shifted copy of the D0-brane tower, and as such its presence does not modify the computation of the species scale. Therefore the asymptotic behaviour of $\Lambda_{\rm sp}$ for the limits $n=3$ is accurate, and that is why the one-loop threshold corrections reproduce this case. 

Things are different for limits with $n=2$, where the assumptions behind \eqref{spD0} no longer hold and one needs to reconsider the computation of the species scale. Indeed, in this case $\Lambda_{\rm sp}$ as it appears in table \ref{tab:scaling_L(phi)} asymptotically decreases much slower than $m_{\rm D2}$ and $\Lambda_{\rm NS5}$. So one should see further towers of states beyond the one of D0's below the UV cut-off $\Lambda_{\rm UV} = \Lambda_{\rm sp}$. In particular, one should consider the tower of multiple D2-branes, whose asymptotics is the same as the D0's. Considering both towers on equal footing one finds that
\be   \label{L_UV_w=2}
\Lambda_{\rm sp} = \frac{M_{\rm P}}{\sqrt{S_{\rm tot}}} = \frac{M_{\rm P}}{\sqrt{S_{\rm D0} S_{\rm D2}}}\, , \quad S_{{\rm D}p} = \frac{\Lambda_{\rm sp}}{\Delta m} \quad \Rightarrow \quad \begin{cases}
\Lambda_{\rm sp} \sim (\Delta m)^{\frac{1}{2}} \sim \phi^{-\frac{1}{2}}\\
S_{{\rm D}p} \sim (\Delta m)^{-\frac{1}{2}} \sim \phi^{\frac{1}{2}}
\end{cases},
\ee
where we have approximated the full tower of states to be a direct product of the D0 and D2 towers, as in \cite{Castellano:2021mmx}. We have also used that $\Delta m \sim \phi^{-1}$ for both of them, which then should have the same $S_{{\rm D}p}$. Notice that this new, corrected scaling of the species scale is similar to $\Lambda_{\rm NS5}$, and so one no longer needs to consider the tower of string states, of which there is at most a finite number below $\Lambda_{\rm UV}$, independently on whether the string is critical or not.

Let us see what this observation implies for the limit $t^2 \to \infty$ in our example. We must now consider the correction to the gauge kinetic functions obtained from integrating out a tower of states of the form 
\be
\textbf{q} = (0,0,0,j,0,-k)\, ,
\label{doubleex}
\ee
where $j$ and $k$ denote the D2$_1$ and D0 charges respectively. The corrections that we get are 
\bes
\label{tildeIex}
\begin{align}
\label{tildeIex00}
\tilde{I}_{00} &\sim \sum_{j,k=-S_{{\rm D}p}}^{S_{{\rm D}p}} k^2 \log \frac{\Lambda_{\rm sp}}{m_{(j,k)}} \sim S_{{\rm D}p} \cdot S_{{\rm D}p}^3 \sim \phi^2\, , \\
\tilde{I}_{11} & \sim \sum_{j,k=-S_{{\rm D}p}}^{S_{{\rm D}p}} j^2 \log \frac{\Lambda_{\rm sp}}{m_{(j,k)}} \sim S_{{\rm D}p} \cdot S_{{\rm D}p}^3 \sim \phi^2\, , \\
\tilde{I}_{22} & \sim \sum_{j,k=-S_{{\rm D}p}}^{S_{{\rm D}p}} 0 \cdot \log \frac{\Lambda_{\rm sp}}{m_{(j,k)}} = 0 \quad \Rightarrow \quad \tilde{I}_{22} \sim \text{const}, \\
\tilde{I}_{12}& \sim \sum_{j,k=-S_{{\rm D}p}}^{S_{{\rm D}p}} j \cdot 0 \cdot \log \frac{\Lambda_{\rm sp}}{m_{(j,k)}} = 0 \quad \Rightarrow \quad \tilde{I}_{12} \sim \text{const},
\end{align}
\ees
which nicely reproduce the appropriate scalings \eqref{Ilim2}.

A comment is in order regarding our choice of charge vector \eqref{doubleex}. A naive application of the results of  \cite{Corvilain:2018lgw} would have lead us to the choice $\textbf{q} = (0,0,0,j,0,-jk)$. Then, when plugged into \eqref{tildeIex00}, this would result into an asymptotic behavior for $\tilde{I}_{00}$ different from the observed value. The fact that \eqref{doubleex} is the most general choice of charge vector can be easily motivated in this case. Indeed, the charge $\textbf{q} = (0,0,0,j,0,0)$ corresponds to a D2-brane wrapping $j$ times the elliptic fibre $\Sigma_2^0$ of this Calabi--Yau. The entry $q_0=k$ of \eqref{doubleex} corresponds to $\int_{\Sigma_2^0} \Tr F$, where $F$ is the the quantised worldvolume flux of the D2-brane. In general, this integral does not need to be a multiple of the multi-wrapping number $j$, but it can be any integer. This is particularly easy to see in toroidal worldvolume geometries like in this case, where a $q_0$ that is not a multiple of $j$ can be achieved by introducing a worldvolume flux of the form $F = \frac{k}{j} \mathbbm{1}_j$ and  non-Abelian Wilson lines that make the gauge bundle well-defined, see e.g. \cite{Guralnik:1997sy,Rabadan:2001mt,Cremades:2004wa}.

Notice also that, given this modified species scale, the tower of D0-branes by itself cannot account for the observed asymptotic behaviour of $\tilde{I}_{00}$. One needs a two-dimensional lattice of charged particles in order to reproduce the behaviour $\phi^2$, with a degenerate spectrum of charges in one of the dimensions. The same  applies to $\tilde{I}_{11}$ and illustrates a general lesson of our analysis: 
\begin{center}
{\em In 4d $\CN=2$ theories the Emergence Proposal requires that the set of charged particles \\ correcting the gauge kinetic terms grows as fast as the number of species.} 
\end{center}
Note that this is a non-trivial requirement, given that in general there can be several towers of particles becoming asymptotically light equally fast, see e.g. \cite{Font:2019cxq}. If they exist, such multi-towers must be of smaller dimension or at most the same dimension as the tower of charged particles, which in this example is of dimension two. When this happens, the lattice of light charged particles effectively sets the species scale. Even if in the example at hand it lowers the naive estimate \eqref{spD0}, there is a compensating effect when summing over the charged lattice. Taking into account the larger degeneration of charges reinstates the expected behaviour $\phi^2$. If there was an additional lattice of uncharged particles with similar asymptotic scale but of dimension $d\geq 3$, then $S_{\rm tot}$ would scale like $\phi^{\frac{2d}{d+2}}$ and such a compensating effect would not occur.

\section{Emergence in $w=2$ limits}
\label{s:w2}

In this section we analyse more general $w=2$ limits, and argue that each of these limits has a double tower of asymptotically massless BPS  particles, including jumps in both D0 and D2-brane charges. As a result, everything works as in the $w=2$ limit analysed in the last section. 

Recall that $w=2$ limits of the form \eqref{limita} are characterised by a vector $e^a \in \mathbbm{N}$ such that ${\bf k} \equiv \CK_{abc}e^ae^be^c =0$ and ${\bf k}_a \equiv \CK_{abc} e^be^c \neq 0$ for some $a$. As emphasised in \cite{Lee:2019wij}, any limit of this kind corresponds to an elliptic fibration over a two-fold base $B_2$, where the self-intersection $e^aD_a \cdot e^aD_a$ describes the class of the elliptic fibre. In the language of \cite{Corvilain:2018lgw}, this corresponds to a limit of  type III$_0$, implying that the matrix ${\bf k}_{ab} \equiv \CK_{abc}e^c$ has rank two. Limits of type III$_c$ with $c>0$ should not be realised in this context. 

 Applying the results of  \cite[Appendix A]{Corvilain:2018lgw}, one finds that the tower of asymptotically massless BPS particles includes the following charges
\be
\textbf{q} = ({\rm D6},\vec{\rm D4}, \vec{\rm D2},{\rm D0}) = \Big(0,\vec{0}, j {\bf k}_a , -  l^a {\bf k}_a \Big),
\label{doublegen}
\ee
 where $\vec{\rm D4}$ and $\vec{\rm D2}$ are vectors of $h^{1,1}(X_6)$ entries, and $j,l^a \in \mathbbm{Z}$. The entries ${\rm D2}_a = j{\bf k}_a$ correspond to the curve class $e^aD_a \cdot e^aD_a$, and one can see that the mass of D2-branes wrapping such a curve asymptote as $\phi^{-1}$ along the limit. That is, \eqref{doublegen} is the generalisation of \eqref{doubleex} for more general $w=2$ limits. Notice that the induced D0-brane charge in \eqref{doublegen} is not necessarily a multiple of $j$. This is an additional input compared to the results of \cite[Appendix A]{Corvilain:2018lgw} that can again be justified by using the geometry of the elliptic fibre.  

To provide a more concrete picture, let us consider smooth elliptic fibrations. Then let us take a K\"ahler cone basis $\omega_a =\{\omega_E, \omega_\a\}$ similar to that in \cite[section 3.4]{Corvilain:2018lgw} where $\omega_\a = \pi^* \omega'_\a$ is the pull-back of a simplicial K\"ahler cone basis of $B_2$. Their triple intersection numbers are
\begin{equation}
    \CK_{EEE} = \eta_{\a\b}c_1^\a c_1^\b, \qquad \CK_{EE\a} = \eta_{\a\b} c_1^\b, \qquad \CK_{E\a\b} = \eta_{\a\b}, \qquad \CK_{\a\b\g} = 0,
    \label{interfib}
\end{equation}
where $c_1(B_2) = c_1^\a \omega'_\a$ and $\eta_{\a\b}$ is a symmetric matrix with signature $(1, h^{1,1}(B_2) -1)$. If we now consider an EFT string limit of the form \eqref{limita} with ${\bf e} = (0, e^\a)$, $e^\a \in \mathbbm{N}$, we find that
\begin{equation}
    {\bf k}_a =  \eta_{\a\b} e^\a e^\b \, \delta_{aE}, \qquad {\bf k}_{ab} =
    \begin{pmatrix}
   \eta_{\a\b} e^\a c_1^\b & \eta_{\a\b} e^{\b} \\  \eta_{\a\b} e^{\b} & 0
    \end{pmatrix} . 
\end{equation}
As advanced, ${\bf k}_{ab}$ has rank 2 and so this corresponds to a limit of type III$_0$. The charge vector \eqref{doublegen} takes the slightly simpler expression 
\begin{equation}
    \textbf{q} =  ({\rm D6},\vec{\rm D4}, \vec{\rm D2},{\rm D0}) = \Big(0,\vec{0}, j \delta_{aE} , -  l^0  \Big)\, ,
\label{vectorell}
\end{equation}
where we have removed an overall integer $\eta_{\bf ee} \equiv \eta_{\a\b} e^\a e^\b>0$. This tower of light states looks like \eqref{doubleex}: we have one particular elliptic curve that generates the D2-brane tower, and on top of it we have the tower of D0-brane induced charges. The set of BPS particles forms a double tower of asymptotically massless states and, if there are no further towers below the species scale, it follows that the number of species $S_{\rm tot}$ and   $\Lambda_{\rm sp}$ also scale as in \eqref{L_UV_w=2}. 

Let us now discuss if there are further towers of states that lie parametrically below the species scale. One may first consider towers of D4-branes wrapping divisors $D_f = f^{\a}D_{\a}$ different from $e^{\a}D_{\a}$. The volume of such divisors is given by 
\begin{equation}
    {\cal V}_{D_f} =  \eta_{\a\b} e^\a f^{\b} \phi + \dots
\end{equation}
where the dots represent terms independent of $\phi$. Therefore, if there was a choice of divisor with $\eta_{\a\b} e^\a f^{\b} =0$, that is a vanishing intersection number with respect to the divisor $e^\a D_\a$ that defines the limit, one could have a tower of light D4-branes. This however cannot happen for divisors that lie in the closure of the K\"ahler cone, namely with $f^\a \in \mathbbm{N}$. At the level of intersections in the two-fold base $B_2$ this can be seen by using the signature of $\eta_{\a\b}$ and the fact that $\eta_{\bf ee} >0$, see Lemma 2 of \cite[Appendix D]{Lee:2019tst}, and at the level of intersections in $X_6$ it follows from Proposition 2 in the same appendix. The remaining effective divisors are contractible four-cycles that should not give rise to a tower of BPS states via multi-wrapping  on them, since their shrinking corresponds to finite-distance singularities in K\"ahler moduli space.\footnote{It remains the possibility to generate a BPS particle tower from a single D4-brane wrapping a contractible divisor, but with different choices of worldvolume flux such that an infinite lattice of induced D2/D0-brane charges is generated. However, from \eqref{interfib} one can check that charges generated in this way must be contained within the lattice \eqref{vectorell}, and so the only effect that this tower could cause is to lower the species scale compared to \eqref{L_UV_w=2}.\label{caveat1}}

Regarding additional towers of D2-branes, they could in principle arise from curves different from the elliptic fibre whose area is independent of $\phi$. However, this can only occur for very specific curves. Notice that those curves that arise from an intersection of Nef divisors of the form $D_E \cdot D_f$ always have an area that depends at least linearly with $\phi$, again because their intersection with the divisor $e^\a D_\a$ cannot vanish. It occurs that those effective curves in the base $B_2$ that do not intersect $e^\a D_\a \cdot B_2$ necessarily have negative self-intersection in $B_2$. As emphasised in \cite{Lee:2019wij}, these are contractible curves whose BPS invariants are non-vanishing for a finite set of multiwrappings, which means that they cannot generate an infinite tower. Finally, one could consider the presence of light D2-branes in elliptic fibrations with singular fibres. This problem is analogous to the analysis of light M2-branes for M-theory compactified on $X_6$ along type $T^2$ limits, performed in \cite[section 3.1]{Lee:2019wij}. There it was found that in order to find asymptotically massless M2-branes wrapping curves of $B_2$ with non-negative self-intersection, it is necessary that $B_2$ admits a fibration structure itself. This case is however incompatible with J-Class A limits, and therefore with $w=2$ EFT string limits. 

To sum up, we find that \eqref{doublegen} represents the dominant lattice of light particles as we proceed along a $w=2$ EFT string limit \eqref{limita}. Following computations similar to those of section \ref{s:simple}, one finds the following asymptotic behaviour for the one-loop corrections \eqref{1loopI}:
\bes
\label{genw2}
\begin{align}
    \tilde{I}_{00} & \sim \phi^2, \\ 
    \tilde{I}_{ab} & \sim \phi^2 \quad \text{if} \quad {\bf k}_{a}\cdot {\bf k}_{b} \neq 0,
\end{align}
\ees
while the remaining corrections vanish. Considering the asymptotic behaviour of the metric \eqref{metric} along this limit
\begin{equation}
    g_{ab} \sim \text{const.}  \quad \text{if} \quad {\bf k}_{a}\cdot {\bf k}_{b} \neq 0,
\end{equation}
and that $\CK \sim \phi^2$, it is easy to see that \eqref{genw2} reproduces the asymptotic behaviour the gauge kinetic functions for all the components that scale like $\phi^2$. 

Notice that the Emergence Principle predicts that the components of $\tilde{I}_{ab}$  either scale like $\phi^2$ or remain constant along a $w=2$ EFT string limit. This is perhaps quite surprising at first, since naively the metric component $g_{ia}$ seems to scale like $\phi^{-1}$ along the limit $t^i = \phi \to \infty$ whenever $\CK_{iia} \neq 0$. If that was the case, the corresponding gauge coupling component would scale like $\tilde{I}_{ia} \sim \phi$. Notice however that a more careful analysis shows that along this limit
\begin{equation}
    \CK \sim 3\phi^2, \qquad \CK_i \sim 2\phi, \qquad \CK_a \sim \phi^2, \qquad \CK_{ia} \sim \phi.
\end{equation}
As a result, we have a cancellation between the two leading terms that appear in $g_{ia}$, so that $g_{ia} \sim \phi^{-2}$ along this limit and $\tilde{I}_{ia}$ remains constant, in agreement with the Emergence Principle.


\section{Emergent EFT string limits}
\label{s:eslim}

In this section we will turn our attention to EFT string limits with scaling weight $w=1$. Recall that in our type IIA setup an EFT string is an NS5-brane wrapped on a Nef divisor specified by the charges $e^a \in \mathbbm{N}$ that implement the axionic shift $b^a \to b^a + e^a$ when circling around the string, as well as the limit \eqref{limita} when we approach the string core along its backreacted solution \cite{Lanza:2021udy}. Recall also that in order for the string modes to behave like the scale $m_*$ of the leading tower of states it must be that ${\bf k}_a \equiv \CK_{abc}e^b e^c = 0$, $\forall a$. 

To evaluate the effect of an EFT string in a limit with scaling weight $w=1$ one must determine {\it i)} if the EFT string corresponds to a dual critical string, in which case we call it an {\em emergent EFT string} and {\it ii)} if the string modes are charged under the $U(1)$'s of the theory. If we have an emergent EFT string, its spectrum will dominate over the towers of light BPS particles made up from wrapped D-branes, whose spectrum we discuss in section \ref{ss:Dtowersw1}, when computing the species scale. If moreover the string oscillations modes are charged, they will dominate the one-loop corrections to the gauge kinetic functions. We will address the charges of the string modes in section \ref{ss:charged}, by applying the anomaly inflow argument of \cite{Heidenreich:2021yda} to our case, finding that they coincide with the D-brane BPS particle charges. Finally, we will consider the effect of all these towers in terms of the Emergence Proposal. The general expectation is that $w=1$ EFT string limits correspond to emergent EFT string regimes. However, since there is no general proof yet, we will   both consider the case in which the string is critical and in which it is non-critical. As advanced, under certain assumptions the asymptotic form of the gauge kinetic function can be obtained in either case.

\subsection{Towers of D-brane particles}
\label{ss:Dtowersw1}

In the classification made in \cite{Lee:2019wij}, EFT string limits with $w=1$ correspond to J-Class B limits with a divisor $D = e^aD_a$ that is either a $K3$ or a $T^4$ fibered over $\mathbbm{P}^1$. The EFT string is made of an NS5-brane wrapped around the fibre, and because ${\bf k}_a \equiv \CK_{abc}e^be^c = 0$ $\forall a$ its volume remains constant along the limit \eqref{limita}. Given this setup, a D4-brane wrapped around $D$ and any D2-brane wrapped on a complex curve within $D$ will have a constant volume along this limit. They will represent BPS particles whose mass only depends on $\phi$ through the 10d dilaton, and therefore display a scaling similar to $m_* = m_{\rm D0}$. 

The corresponding lattice of charges can be represented as 
\be
\textbf{q} = ({\rm D6},\vec{\rm D4}, \vec{\rm D2},{\rm D0}) = \Big(0,q_{\rm D4} e^a, {\bf k}_{ab}w^b, q_{\rm D0} \Big) \cdot Q^{t\, -1}, \qquad q_{\rm D0}, \, q_{\rm D4}, w^b \in \mathbbm{Z},
\label{multigen}
\ee
where ${\bf k}_{ab} \equiv \CK_{abc}e^c$ and $Q$ is given by \eqref{PaQ}. Here $q_{\rm D4}$ and $q_{\rm D0}$ represent the D4 and D0-brane charges, while $Q$ implements the lower-dimensional charges induced by the curvature when a D4-brane is wrapped around $D$. The D2-brane charges ${\rm D2}_a = {\bf k}_{ab}w^b$ correspond to the monodromy seed vectors found in \cite[Appendix A]{Corvilain:2018lgw} for II$_b$ limits, with $b$ is the rank of ${\bf k}_{ab}$. When $w^a \in \mathbbm{N}$ $\forall a$, one can interpret these charges as those curves within $D$ that arise from the intersection with other Nef divisors. More generally, one can see it as a sublattice of the Picard lattice $ {\rm Pic}(D) = H^{1,1}(D) \cap H^2(D, \mathbbm{Z})$. For D2-branes to be BPS objects they need to wrap non-trivial curves in $X_6$, which means that they should be seen as elements of the lattice dual to $\Lambda^\vee = \iota^* H^2(X_6, \mathbbm{Z})$, where $\iota$ is the embedding map of $K3$ into $X_6$. As pointed out in \cite{Lee:2019wij} for these BPS states to form a tower we must also require that they wrap a curve with non-negative self-intersection number in $D$, which in the case of generic $K3$ fibres results into a one-dimensional lattice of states. 

While these observations restrict the spectrum of BPS states with pure D2-brane charge, notice that  there is a different way to generate a vector with D2-brane charges, namely by considering a quantised field strength $F$ on top of the D4-brane wrapping $D$. Such charges are not constrained by the self-intersection condition and dominate the charge spectrum in \eqref{multigen}, because for pure D2-brane charges we need to impose that $q_{\rm D4} =0$. Such quantised fluxes are elements of ${\rm Pic} (D)$, but their $U(1)$ charges belong to the sublattice $\Lambda^\vee$, and the elements of the latter (or that of a finite-index sublattice) are represented by the entries ${\bf k}_{ab}w^b$ with $w^b \in \mathbbm{Z}$. Taking all this into account, one can see that the lattice of $U(1)$ charges populated by asymptotically massless BPS states has dimension $2 +b$, where $b = {\rm rank}\, {\bf k}_{ab}$ is the subindex in II$_b$ that defines the subtype of limit in the classification of \cite{Grimm:2018ohb,Grimm:2018cpv,Corvilain:2018lgw}. We find quite interesting that this subindex, obtained from a purely mathematical classification of infinite distance limits, has a direct physical meaning in terms of the lattice of asymptotically massless D-brane  states.\footnote{Notice that $b$ only fixes the dimension of the lattice of D-brane particle charges. The dimension of the lattice itself should instead correspond to $2 + \dim {\rm Pic}(D)$.\label{caveat2}}

Finally, notice that in \eqref{multigen} we have assumed that D2 and D0-brane charges scan integer values independently of $q_{\rm D4}$, in analogy with the assumptions in $w=2$ limits. This assumption is again valid for $w=1$ limits where $D$ is a $T^4$, while for $K3$ fibres it can be motivated by considering D-brane charges at an orbifold limit of $K3$.

\subsection{Charged string modes}
\label{ss:charged}

To determine the $U(1)$ charges of the oscillations of an EFT string one may follow the strategy outlined in \cite[Appendix A]{Heidenreich:2021yda}. In there, the need of localised degrees of freedom on the axionic string worldsheet is made manifest via an inconsistency of the equations of motion for the gauge fields, which is then fixed when charged string modes are included. 

To detect the analogous inconsistencies in our setup, let us recall from section \ref{s:scaling} that the entries of the vector ${\bf {F}}^{\, t} = (F^0, F^a, F_{a} ,-F_{0})$ are not invariant under axionic shifts. Instead, they are the entries of $\bar{\cR} {\bf F}$, with $\bar{\cR}$ given by \eqref{monoexp}, that are invariant. As such, they should be globally well-defined quantities in any configuration, including a backreacted EFT string solution. This in turn implies that
\be
0 = d^2 (\bar{\cR} {\bf F}) = \bar{\cR} (d^2 {\bf F}) + (d^2 \bar{\cR}) {\bf F}\, ,
\ee
which leads to
\be
d^2{\bf F} = - \bar{\cR}^{-1} (d^2 \bar{\cR}) {\bf F} = - d^2 (b^a) \wedge \bar{P}_a {\bf F}\, ,
\ee
where we have used $\bar{\cR}=Q e^{b^a \bar{P}_a}$. In the absence of axionic strings  $d^2 (b^a) = 0$, and we recover that $d^2{\bf F} = 0$, as expected from the usual bulk Bianchi identities $dF^A =0$ and equations of motion $dF_A =0$, see Appendix \ref{ap:democratic}. In the presence of an axionic string with charge $e^a$ we instead have that $d^2 (b^a) = e^a \delta(\Sigma)$, where $\Sigma$ is the world-sheet of the string. Using the expression \eqref{PaQ} for $\bar{P}_a$ we then obtain
\be
 \begin{pmatrix}
d^2 F^0\\
d^2 F^a\\
d^2 \tilde{F}_{a}\\
- d^2 \tilde{F}_{0}
\end{pmatrix} = - 
\begin{pmatrix}
0\\
e^a F^0\\
{\bf k}_{ab} F^b \\
- e^a \tilde{F}_a 
\end{pmatrix} \wedge \delta({\Sigma})\, ,
\label{d2F}
\ee
where we have defined $\tilde{\bf F} = Q \cdot {\bf F}$. This reflects the fact that the entries of {\bf F} are not invariant under axionic shifts, with the exception of $F^0$. 

From the viewpoint of \cite{Heidenreich:2021yda} $d^2F_A \neq 0$ signals sources for the gauge fields equations of motion, that correspond to charged string degrees of freedom. More precisely one introduces the vector
\begin{equation}
    \Xi = \begin{pmatrix}
0\\
\xi^a \\
\xi_a \\
\xi_0
\end{pmatrix}\wedge \delta({\Sigma})\, ,
\end{equation}
of terms localised on the axionic string worldsheet, such that 
\be
d F^a = \xi^a \wedge \delta(\Sigma)\, ,  \qquad d \tilde{F}_a = \xi_a \wedge \delta(\Sigma)\, , \qquad
d\tilde{F}_0 = \xi_0 \wedge \delta(\Sigma)\, ,  
\ee
provide the sources needed to reproduce \eqref{d2F}. Finally, these can be solved by taking 
\be
\xi^a = - e^a A^0 + d\eta^a\, , 
\qquad \xi_a = - {\bf k}_{ab} e^b A^c + d\eta_a\, , 
\qquad 
\xi_0 = - e^a \tilde{A}_a  + d\eta_0\, . 
\ee
Following \cite[Appendix A]{Heidenreich:2021yda}, these  indicate the charges of localised string modes under different gauge transformations
\bes
\label{locgauge}
\begin{align}
\label{locgaugeD0}
A^0 \mapsto A^0 + d\lambda^0, & \qquad \eta^a \mapsto \eta^a +  e^a \lambda^0\, , \\
\label{locgaugeD2}
A^a \mapsto A^a + d\lambda^a, & \qquad \eta_a \mapsto \eta_a + \cK_{abc} e^b \lambda^c\, , \\
\label{locgaugeD4}
\tilde{A}_a \mapsto \tilde{A}_a + d\lambda_a, & \qquad \eta_0 \mapsto \eta_0 + e^a \lambda_a\, .
\end{align}
\ees
Remarkably, these charges match the spectrum of BPS particle charges found in \eqref{multigen}, that are asymptotically massless along the EFT string limit. Following \cite{Lee:2019wij}, one expects that in emergent string limits all D-brane particle states map to the BPS spectrum of a dual critical string. EFT string oscillation modes and bound states of both should complete the spectrum of such a critical string. It is then natural that EFT string oscillations have the same charges as the D-brane particles, if in general they correspond to bound states of uncharged oscillations and BPS states. If such bound states are moreover stable, they will dominate the light string spectrum. As we will see below, this is required in order to reproduce the asymptotic behaviour of the gauge kinetic functions via the Emergent Principle in emergent EFT string limits.

\subsection{Emergence in $w=1$ limits}
\label{ss:stower}

Let us now consider the computation of the species scale in $w=1$ limits. As can be appreciated from table \ref{tab:scaling_L(phi)}, in this kind of limits there are three different kinds of towers that scale like $m_* \sim \phi^{-\frac{1}{2}}$: the tower of Calabi--Yau Kaluza--Klein modes, the BPS tower of D-brane particles (including D4-branes wrapped on $D$), and the tower of EFT string oscillations. Following this picture, one can estimate the species scale to be of the form 
\be{}   \label{L_UV_w1}
\Lambda_{\rm sp} = \frac{M_{\rm P}}{\sqrt{S_{\rm tot}}} \simeq \frac{M_{\rm P}}{\sqrt{S_{\rm KK} + S_{{\rm D}p} + S_{\text{str}}}},
\ee{}
where $S_{\rm KK}$ represent the tower of KK modes along the $\mathbbm{P}^1$ base of $X_6$, $S_{{\rm D}p}$ the BPS D-brane particles and $S_{\rm str}$ the EFT string oscillation modes. If these sets form bound states then this expression should be modified, although as we will see this will not make much difference.  

What is more relevant for the spectrum of light particles is whether or not the EFT string is dual to a critical string. If the string is non-critical it will have a finite number of oscillation modes. Even if these form bound states with the D-brane states, this will only result in having a finite number of copies of the BPS D-brane particle spectrum, and so one can essentially remove $S_{\rm str}$ from \eqref{L_UV_w1}. In addition D-brane states will dominate over the CY KK modes, so in practice one can write $\Lambda_{\rm sp} = M_{\rm P}/\sqrt{ S_{{\rm D}p}}$. If the dual string is critical, then  D-brane states and EFT oscillations will combine into a single spectrum of such a dual string, which will again dominate over the KK sector. We thus end up with an expression of the form $\Lambda_{\rm sp} = M_{\rm P}/\sqrt{ S_{\rm crit}}$.

As already mentioned, the expectation is that all $w=1$ EFT string limits correspond to emergent string limits, or in other words to a critical string.  While the results of \cite{Lee:2019wij} support this expectation, they do not provide a complete proof. Indeed, recall that in EFT string limits the divisor $D$ fibered over $\mathbbm{P}^1$ remains of constant volume as we move in moduli space. This is captured by J-Class B limits where $D$ is a $K3$, because even if classically one shrinks the $K3$ fibre, there is a quantum obstruction to do it after a certain point, after which one can only grow the base. Such kinds of limits, with the dilaton co-scaling \eqref{codila}, have scaling weight $w=1$ and are emergent EFT string limits, with an emergent heterotic string \cite{Lee:2019wij}. 

The situation is less clear for J-Class B limits with a $T^4$ fibre. In that case, limits with a scaling like in \cite[eq.(4.73)]{Lee:2019wij} are not EFT string limits of the form \eqref{limita}. One can actually check that \eqref{scaling} is satisfied with $w=2$, which agrees with the result that these are decompactification limits. To convert such a limit into an EFT string limit one should do an overall rescaling of the K\"ahler moduli, which will change the nature of the limit and in particular the scaling weight to $w=1$. It is expected\footnote{We would like to thank Timo Weigand for discussion about this point.} that such a limit behaves like J-Class B limits of M-theory with $T^4$ fibre. These were argued in \cite{Lee:2019wij} to lead to an emergent string limit with a dual type IIB critical string probing a non-geometric D-manifold, based on the case in which the fibre factorises as $T^2 \times T^2$. 

While it seems likely that in our setup EFT string limits with $w=1$ are emergent EFT string limits, we will take a general approach and consider two different scenarios. We will first assume the case where the EFT string is non-critical, since then the computations are relatively similar to those for $w=2$ limits. We will then discuss the more involved case of a critical EFT string, assuming a spectrum dominated by charged particles. Remarkably, one is able to reproduce the expected asymptotic behaviour for the gauge kinetic function in both scenarios. Thus, if our assumptions are correct, it seems that as far as the Emergence Proposal is concerned one is not preferred over the other.

\subsubsection*{Non-critical EFT string}

If the EFT string is non-critical, $S_{\rm str}$ contains a finite number of elements, and for computing the species scale it can be neglected. KK modes also form an infinite tower, but it is always smaller in dimension than the tower of D-brane particle states. 
Therefore in the following we will ignore the presence of $S_{\rm KK}$ in \eqref{L_UV_w1}. Also, to avoid the subtleties of footnotes \ref{caveat1} and \ref{caveat2}, we will take the simplifying assumption that we are in a II$_b$ limit such that $b=\dim {\rm Pic}(D)$. Relaxing these conditions will not change the final outcome. 

With these assumptions in hand and the same approximations as in section \ref{s:simple} we have that
\be{}   \label{L_UV_w1_nc}
\Lambda_{\rm sp} = \frac{M_{\rm P}}{\sqrt{S_{\rm tot}}} = \frac{M_{\rm P}}{ \sqrt{ S_{{\rm D}4} \cdot S_{{\rm D}2}^b \cdot  S_{{\rm D}0}}}, \qquad S_{{\rm D}p} = \frac{ \Lambda_{\rm sp}}{\Delta m} ,
\ee{}
from where we get the following scalings
\be
\Lambda_{\rm sp} \sim m_*^{\frac{2+b}{4+b}} \sim \phi^{-\frac{2+b}{2(4+b)}}, \qquad S_{{\rm D}p} \sim \frac{\Lambda_{\rm sp}}{\Delta m} \sim \phi^{\frac{1}{4+b}}. 
\ee
The one-loop corrections to gauge kinetic functions can then be computed following similar calculations to those in section \ref{s:simple}. For instance, in the case of the graviphoton we have that the corrections to $\tilde{I}_{00}$ read
\be
\sum_{{\bf q},\, |q_i| \leq S_{{\rm D}p} } q_{\rm D0}^2 \log \frac{\Lambda_{\rm sp}}{m_{{\bf q}}} \sim
S_{{\rm D}p}^{b+1} \cdot S_{{\rm D}p}^3 \sim \phi .
\label{estDpart00}
\ee
Notice that in the above computations it is important to assume that $q_{\rm D0}$ in \eqref{multigen} is independent from $q_{\rm D4}$ and $w^a$, so that the sum over $q_{\rm D0}$ decouples from the other components of the charge vector {\bf q} when computing the corrections to $\tilde{I}_{00}$. This decoupling will in general not occur for D2-brane charges, which  contribute to a correction to the component $\tilde{I}_{ab}$ of the form
\be
\sum_{{\bf q}(w^a),\, |q_i| \leq S_{\rm Dp} } {\bf k}_{ac} w^c {\bf k}_{bd} w^d\,  \log \frac{\Lambda_{\rm sp}}{m_{{\bf q}}} . 
\label{estDpartab}
\ee
If the matrix ${\bf k}_{ab}$ was diagonal, each sum would decouple and, just like in \eqref{estDpart00}, we would recover a correction of the form $S_{{\rm D}p}^{b+4} \sim \phi$ for each component $\tilde{I}_{aa}$ such that ${\bf k}_{aa} \neq 0$. In general it is not true that ${\bf k}_{ab}$ is diagonal, but by looking at explicit examples like the ones in the next section, one can convince oneself the corrections still scale like $\phi$. More precisely, one finds that corrections to $\tilde{I}_{ab}$ scale like $\phi$ if the sum \eqref{estDpartab} is non-trivial, but that several of these corrections are related to each other if the value of $b = {\rm rank}\, {\bf k}_{ab}$ is not maximal. While it would be interesting to derive a more precise relation, one can see that this picture reproduces the asymptotic behaviour of $\tilde{I}_{ab} = \frac{2}{3} \CK g_{ab}$ along these limits. 

Indeed, in II$_b$ limits of the form $t^a = e^a \phi$, $\phi \to \infty$  there are three different kinds of field directions $v^a$ with respect to the behaviour of the entries of $\tilde{I}_{ab}$:
\begin{equation}    \label{field_dir}
   {i)}\ v_{\rm I}^a \notin \ker {\bf k}_{ab}, \qquad {ii)}\ v_{\rm II}^a \in  \ker {\bf k}_{ab} \ \text{and}\ v_{\rm II}^a \perp e^a, \qquad {iii)}\  e^a .
\end{equation}
In terms of these directions we have the following asymptotic behaviour for $\tilde{I}_{ab}$ components:
\bes
\label{estasympto}
\begin{align}
    \tilde{I}_{ab}  v_{\rm I}^a w_{\rm I}^b  & \sim \phi\, ,\\
    \tilde{I}_{ab}  v_{\rm II}^a  w_{\rm II}^b, \quad  \tilde{I}_{ab}  v_{\rm II}^a  w_{\rm I}^b,  &\sim \text{const}\, , \\
     \tilde{I}_{ab} e^a w^b  & \sim \phi^{-1}\, ,
     \label{estasympto3}
\end{align}
\ees
where in the last line $w^b$ is an arbitrary vector. As expected, the first and second cases are easy to understand from the viewpoint of the Emergence Proposal, as they correspond to directions in which the sum \eqref{estDpartab} is non-trivial and trivial, respectively. 

Even the  vanishing behaviour \eqref{estasympto3} has a neat interpretation in terms of the Emergence Proposal. Indeed, an asymptotic behaviour of the form $\phi^{-1}$ for some components of $\tilde{I}_{ab}$ can be seen as a divergence of the form $\phi$ for the magnetic dual $\tilde{I}^{ab} = \frac{3}{2} \CK^{-1} g^{ab}$. For this divergence to occur, one should have a tower of magnetically charged particles contributing to the one-loop threshold corrections. This is precisely what happens in this kind of limits, in which an asymptotically massless tower of D4-branes corrects the gauge kinetic functions, with a charge along the magnetically divergent direction $e^a$. While from the viewpoint of $\tilde{I}_{ab}$ there are many mixed components besides $\tilde{I}_{ab}e^ae^b$ that decrease like $\phi^{-1}$, one can see that this is a result of expressing such magnetic divergences in the electric basis. As can be verified in the examples of the next section, in terms of the dual kinetic matrix  $\tilde{I}^{ab}$ there is a single component that diverges, which is precisely the one that couples to the D4-brane charge. In this dual framework, the threshold corrections for such a component read as in \eqref{estDpart00} with the replacement $q_{\rm D0} \to q_{\rm D4}$, yielding the expected linear behaviour. Thus, even those $U(1)$ gauge theories that run towards strong coupling in the boundary of the moduli space can be understood as an IR effect. 

\subsubsection*{Critical EFT string}

We now consider the case in which the D-brane particles and EFT string modes combine into the spectrum of a dual critical string. In this case 
the string contains an infinite number of oscillation modes, and the number of states in the string tower is expected to grow exponentially with the mass level $N$, with a mass spectrum of the form $m_N^2 \sim \cT N$, see \cite{Lee:2018urn,Heidenreich:2017sim} for examples. The species scale in this case reads
\be \label{cstr_sp}
\Lambda_{\rm sp}^2 = \frac{M_{\rm P}^2}{S_{\rm crit}}\, , \qquad S_{\rm crit} = \sum_{N=0}^{N_{\rm max}} S_{\rm crit}^{(N)}\, ,
\ee
where $S_{\rm crit}^{(N)}$ is the number of states at mass level $N$ and $N_{\max}$ corresponds to states whose mass lies at the species scale $\cT N_{\rm max} \sim \Lambda_{\rm sp}^2$. Since an exponential degeneracy is expected, and following the approach applied in \cite{Castellano:2022bvr} to other setups, we make the following Ansatz
\be \label{cstr_deg}
S_{\rm crit}^{(N)} \sim e^{N^\alpha}N^{\gamma}\, , \qquad \alpha>0\, , \gamma \in \mathbbm{R}\, .
\ee
In the analysis of \cite{Castellano:2022bvr}, the parameters of this Ansatz are fixed to $\alpha=\frac{1}{2}$ and $\gamma$ to its value in each 10d string theory case, assuming that the 10d behaviour persists in lower dimensions. Here we choose to keep them general, in order to understand better what are the key assumptions needed for the Emergence Proposal to work. The number of species, at leading order, reads
\be
S_{\rm crit} \sim e^{N_{\rm max}^\alpha} N_{\rm max}^{\gamma+1-\alpha}\, ,
\ee
which when substituted in \eqref{cstr_sp} gives the following scalings
\be
\label{scalingsw1}
\begin{split}
&N_{\rm max} \sim \left[ \log \frac{M_{\rm P}^2}{\cT} -  \left(\frac{\gamma+2}{\alpha}-1\right)\log \left( \log \frac{M_{\rm P}^2}{\cT} \right) \right]^{\frac{1}{\alpha}} , \\
&\Lambda_{\rm sp} \sim \sqrt{\cT} \left[ \log \frac{M_{\rm P}^2}{\cT} -  \left(\frac{\gamma+2}{\alpha}-1\right)\log \left( \log \frac{M_{\rm P}^2}{\cT} \right) \right]^{\frac{1}{2\alpha}} ,\\
&S_{\rm crit} \sim \frac{1}{N_{\rm max}} \frac{M_P^2}{\cT} \sim \frac{M_{\rm P}^2}{\cT} \left[ \log \frac{M_{\rm P}^2}{\cT} - \left(\frac{\gamma+2}{\alpha}-1\right)\log \left( \log \frac{M_{\rm P}^2}{\cT} \right) \right]^{-\frac{1}{\alpha}} .
\end{split}
\ee
Additionally, it has been found in \cite{Castellano:2022bvr} that when one computes the 1-loop correction to the exact gauge field propagator given by any particle in the tower, \eqref{1loopI} must be modified by substituting
\begin{equation}
    \log \left( \frac{\Lambda_{\rm sp}}{m_{N}} \right) \quad \to \quad \log \left( \frac{\Lambda_{\rm sp}}{m_{N}} +c \right), 
\end{equation}
with $c \in \mathbbm{R}_{>0}$ some constant. This change does not affect the outcome of threshold corrections in $w=3$, $w=2$ and $w=1$ non-critical  limits, but it does make a difference in the present setup. 

At this point, in order to compute the 1-loop corrections to the gauge kinetic function, we need to make an Ansatz for the spectrum of $U(1)$ charges of the string tower. Let us first consider the toy example of a single charge.  We assume to have, at each mass level $N$, a maximal charge that scales like $q_{\rm max}(N) \sim \sqrt{N}$. This is the expectation for a critical heterotic string (see e.g. \cite{Lee:2018urn}), and in the present context modes with $q= q_{\rm max}$ can be interpreted as those states that saturate the BPS bound. We denote by $f_N(q)$ the number of states with mass $m_N$ and charge $q$, with $f_N(q)$ a general function which vanishes for $q>q_{\rm max}(N)$. Motivated by the general lesson of section \ref{s:simple}, namely that charged particles must dominate the spectrum, we constrain this function by imposing that 
\be
\sum_{q=0}^{q_{\rm max}(N)} f_N(q) \sim e^{N^\a} N^\gamma\, ,
\label{assumptq}
\ee
so that, by summing over all the charges at each mass level, we essentially get the level degeneracy \eqref{cstr_deg}. 
 In this case, one easily recovers the expected correction to the gauge kinetic function, independently on the value of the parameters $\alpha$ and $\gamma$:  
\be \label{phi_scal1}
\begin{split}
\tilde{I} &\sim \sum_{N=1}^{N_{\rm max}} \sum_{q=1}^{q_{\rm max}(N)} f_N(q) \, q^2 \log \left( \frac{\Lambda_{\rm sp}}{m_{N}} +c \right) \sim \sum_{N=1}^{N_{\rm max}} e^{N^\alpha} N^{\gamma+ 1} \log \left( \frac{N_{\rm max}}{N} +c\right)\\
&\sim e^{N_{\rm max}^\alpha} N_{\rm max}^{\gamma+2-\alpha} \sim \frac{M_{\rm P}^2}{\cT} \sim \phi\, ,
\end{split}
\ee
where in the second line we have approximated $\sum_{n=1}^N f(n) \log \left( \frac{N}{n} +c \right) \sim g(N) \log(1+c) \sim g(N)$, with $g'=f$. Notice that if we set $c=0$ the leading term in \eqref{phi_scal1} would be different, since the logarithm would suppress the corrections from the heaviest states in the tower. Instead of the correct asymptotic behaviour,  we would then obtain that  $\tilde{I} \sim \phi \log \phi$.

One can easily adapt this Ansatz to an arbitrary number of $U(1)$ gauge fields. We again have \eqref{cstr_deg} and the corresponding scalings \eqref{scalingsw1}, as well as the maximal charge scaling  $q_{\rm max}(N) \sim \sqrt{N}$. The spectrum of charges is now indexed by a vector $\vec{q}=(q_0,...,q_{b+1})$ made of independent entries of {\bf q} in \eqref{multigen}. In terms of this new vector we consider a generic function $f_N(q_j)$ for the number of states with mass $m_N$ and charges $\vec{q}=(q_j)$, which is vanishing if $|\vec{q}|>q_{\rm max}(N)$ and constrained in such a way that
\be \label{cons_str2}
\sum_{|\vec{q}|<q_{\rm max}(N)} f_N(q_j) \sim e^{N^\alpha} N^{\gamma}\, .
\ee
Note that in \eqref{cons_str2}, if we approximate the sphere of radius $q_{\rm max}(N)$ in the space of charge vectors $\vec{q}$ with a hyper-cube of side $q_{\rm max}(N)$, we just change the result by a multiplicative factor, as in \cite{Castellano:2021mmx}. Finally, we compute the corrections to the diagonal entries of the gauge kinetic function
\bea \nonumber
\tilde{I}_{AA} &\sim& \sum_{N=1}^{N_{\rm max}} \sum_{|\vec{q}|<q_{\rm max}(N)} f_N(q_j) \, q_A^2 \left( \log \frac{\Lambda_{\rm sp}}{m_{\vec{q}}} +c \right) \sim \sum_{N=1}^{N_{\rm max}} e^{N^\alpha} N^{\gamma+1} \log \left( \frac{N_{\rm max}}{N} +c \right)\\
&\sim& e^{N_{\rm max}^\alpha} N_{\rm max}^{\gamma+1-2\alpha} \sim \frac{M_{\rm P}^2}{\cT} \sim \phi\, ,
\label{phi_scal2}
\eea
where we have taken the sphere $\to$ cube approximation  in such a way that the sums over the different components $q_j$ factorise. Again, the result is independent on the values of the parameters $\alpha$ and $\gamma$, and the scaling matches with the desired behaviour.

Notice that we have made two main assumptions to arrive to this result:
\begin{enumerate}
    \item  We have a degeneracy spectrum of the form \eqref{cstr_deg}, that generalises the standard critical string spectrum.
    
    \item Charged modes either dominate or are a significant fraction of the critical string spectrum, which translates into the relation \eqref{assumptq}.

\end{enumerate}
The second of these assumptions follows the lesson learnt in section \ref{s:simple}, and seems unavoidable for the Emergence Proposal to work in $\CN=2$ setups, since only charged states can generate corrections for the gauge kinetic function and the moduli space metric at the same time. If the  corrections to the metric and the gauge kinetic function were uncorrelated it would be difficult to reproduce the standard $\CN=2$ supergravity relations. In our setup, this hypothesis could be justified in a number of ways, for instance if EFT string oscillation modes form stable bound states with D-brane particles. In fact, from the viewpoint of $\CN=2$ theories the most natural option is that BPS states lead the one-loop corrections to the gauge couplings and metric, because then their said relations are automatically preserved. However, it would remain to understand the microscopic realisation of such an enhanced BPS spectrum, either in the present type IIA setup or from the dual emergent string viewpoint.


\section{Further examples}
\label{s:examples}

In this section we present two further examples of type IIA CY  compactifications and discuss how emergence works for them. In particular, we consider a two-K\"ahler moduli and a three-K\"ahler moduli example, both containing decompactification as well as emergent string limits.

\subsubsection*{Two-moduli example}

We start with the two-moduli example, which was first presented in \cite{Candelas:1993dm,Hosono:1993qy}, and recently studied in \cite{Lee:2019wij}. In this model the CY is given by the projective space $\mathbb{P}^5_{1,1,2,2,6}[12]$ and is a K3 fibration over a $\mathbb{P}^1$ base. We denote the K\"ahler moduli by $t^a,a=1,2$, the triple intersection numbers are
\be \label{tripint_ex2}
\CK_{111}=4, \quad \CK_{112}=2, \quad \CK_{122}=0, \quad \CK_{222}=0\, ,
\ee
and there are two elementary infinite distance limits:
\begin{enumerate}
\item $t^1 \rightarrow \infty \qquad w=3$\, ,
\item $t^2 \rightarrow \infty, \qquad w=1$\, .
\end{enumerate}
We are interested in the scaling of the gauge kinetic matrix, whose expression is given by \eqref{gaugekin} along each of these limits. In the $t^1 \rightarrow \infty$ limit we have $\CK \sim \phi^3$ and the components of the moduli space metric all scale like $g_{ab} \sim \phi^{-2}$, hence
\be
\tilde{I}_{00} \sim \phi^3, \quad \tilde{I}_{ab} \sim \phi\, .
\ee
These scalings are the ones expected for a $w=3$ limit, as shown in table \ref{tab:limits}, and as we have seen they are reproduced by integrating out a tower of D2-D0 bound states, with fixed D2 and increasing D0 charge, which represents the lightest tower along such type of limits, see table \ref{tab:scaling_L(phi)}.

In the $t^2 \rightarrow \infty$ limit instead we find $\CK \sim \phi$ and $g_{11} \sim \text{const}$, $g_{12} \sim g_{22} \sim \phi^{-2}$, which imply
\be
\tilde{I}_{00} \sim \phi, \quad \tilde{I}_{11} \sim \phi, \quad \tilde{I}_{12} \sim \tilde{I}_{22} \sim \phi^{-1}\, .
\ee
Here we are in a $w=1$ limit, namely an emergent string limit, so in addition to the towers of D$p$-particles we need to consider the tower of string oscillations, which are becoming light with the same rate, see table \ref{tab:scaling_L(phi)}. As regards the components $\tilde{I}_{00},\tilde{I}_{11}$, we have shown in the previous section that one can recover the $\phi$ scaling with a particular Ansatz for the spectrum of the string oscillations tower, as in \eqref{phi_scal1} and \eqref{phi_scal2}. Furthermore, using the anomaly inflow argument of section \ref{ss:charged}, in particular equations \eqref{locgauge}, one can argue for the fact that these are the only components that receive {\it electric} corrections from the string tower. In particular, substituting $e^a=\delta^a_2$ in these equations, \eqref{locgaugeD0} shows that the localised modes are charged under the graviphoton and \eqref{locgaugeD2}, together with \eqref{tripint_ex2}, shows that they are charged under $A^1$. As explained in the last section, the meaning of the other components of the gauge kinetic matrix is more subtle. To explain it, we first compute the scaling of the inverse gauge kinetic function
\be
\tilde{I} = -\frac{\CK}{6}
\begin{pmatrix}
1 & 0\\
0 & 4g_{ab}
\end{pmatrix}\, , \qquad
\tilde{I}^{-1} = -\frac{6}{\CK}
\begin{pmatrix}
1 & 0\\
0 & \frac{1}{4}g^{ab}
\end{pmatrix}\, ,
\ee
where at leading order we have
\be
g^{ab} = \begin{pmatrix}
2(t^1)^2 & -\frac{4}{3}(t^1)^2\\
-\frac{4}{3}(t^1)^2 & 4\phi^2
\end{pmatrix}\, .
\ee
The inverse of the kinetic matrix corresponds to the gauge couplings in the magnetic dual frame. The only magnetically charged states that become asymptotically massless are D4-brane wrapping the divisor $e^a D_a = D_2$ of the EFT string. Via the one-loop corrections, they source the magnetic dual of the gauge field $U(2)_2$, which coincides with the fact that the  only diverging component of $g^{ab}$ is $g^{22}$.

\subsubsection*{Three-moduli example}

The second example contains three vector multiplets, it was studied in e.g. \cite{Scheidegger:2001cqa,Lee:2019wij}, and in this case the CY three-fold is $\mathbb{P}^5_{1,1,2,8,12}[24]$, which admits both a K3 and a $T^2$ fibrations. We denote the K\"ahler moduli as $t^a, a=1,2,3$ and the triple intersection numbers read
\be \label{tripint_ex3}
\CK_{122}=2, \quad \CK_{123}=1, \quad \CK_{222}=8, \quad \CK_{223}=4, \quad \CK_{233}=2\, .
\ee
There are three elementary infinite distance limits
\begin{enumerate}
\item $t^1 \rightarrow \infty \qquad w=1$\, ,
\item $t^2 \rightarrow \infty \qquad w=3$\, ,
\item $t^3 \rightarrow \infty \qquad w=2$\, .
\end{enumerate}
The $t^1 \rightarrow \infty$ limit is an emergent string limit and we find $\CK \sim \phi$, and $g_{22} \sim g_{23} \sim g_{33} \sim \text{const}$, $g_{11} \sim g_{12} \sim g_{13} \sim \phi^{-2}$, which imply
\be \label{ex3_w=1}
\tilde{I}_{00} \sim \phi, \qquad \tilde{I}_{22} \sim \tilde{I}_{23} \sim \tilde{I}_{33} \sim \phi, \qquad \tilde{I}_{11} \sim \tilde{I}_{12} \sim \tilde{I}_{13} \sim \phi^{-1}\, .
\ee
As in the previous example, we recover the scaling of the diverging components of the gauge kinetic matrix by integrating out the tower of EFT string states. In particular, using \eqref{locgauge} and substituting $e^a=\delta^a_1$ and \eqref{tripint_ex3}, one can see that the localised degrees of freedom on the world-sheet carry electric charge under the graviphoton, and the fields $A^2,A^3$, precisely the ones with corrections proportional to $\phi$. Also in this case, we can make sense of the scaling of the other components by going to the magnetic dual frame and computing the inverse kinetic function. In this case we have that $g^{11} \sim 4\phi^2$, while all the other components of $g^{ab}$ are constant. 

Let us also comment on how the classification \eqref{field_dir} of field directions is realised in this limit. The matrix ${\bf k}$, with components ${\bf k}_{ab} = \CK_{1ab}$ takes the form
\be
{\bf k}_{ab} = \begin{pmatrix}
0 & 0 & 0\\
0 & 2 & 1\\
0 & 1 & 0
\end{pmatrix}\, ,
\ee
and we have
\begin{equation}
   {i)}\ v_{\rm I}^a \in \left\langle \begin{pmatrix} 0\\1\\0 \end{pmatrix},
   \begin{pmatrix} 0\\0\\1 \end{pmatrix} \right\rangle\, , \qquad
   {ii)}\ v_{\rm II}^a =0\, , \qquad {iii)}\  e^a = \begin{pmatrix} 1\\0\\0 \end{pmatrix}\, .
\end{equation}
It is then easy to check that the scalings \eqref{ex3_w=1} match with \eqref{estasympto}.

In the $t^2 \rightarrow \infty$ limit we have the scalings $\CK \sim \phi^3$ and $g_{ab} \sim \phi^{-2}$, and then
\be
\tilde{I}_{00} \sim \phi^3, \qquad \tilde{I}_{ab} \sim \phi\, ,
\ee
just like in the previous $w=3$ limits.

Finally, in the $t^3 \rightarrow \infty$ limit we get $\CK \sim \phi^2$ and $g_{22} \sim \text{const}$, $g_{11} \sim g_{12} \sim g_{13} \sim g_{23} \sim g_{33} \sim \phi^{-2}$, and hence
\be \label{ex3_w=2}
\tilde{I}_{00} \sim \phi^2, \qquad \tilde{I}_{22} \sim \phi^2, \qquad \tilde{I}_{11} \sim \tilde{I}_{12} \sim \tilde{I}_{13} \sim \tilde{I}_{23} \sim \tilde{I}_{33} \sim \text{const}\, .
\ee
Let us also match this example with the more general discussion of section \ref{s:w2}. In this case $e^a=\delta^a_2$ and the matrix ${\bf k}_{ab}$ and the vector ${\bf k}_a$ read
\be
{\bf k}_{ab} = \begin{pmatrix}
0 & 1 & 0\\
1 & 4 & 2\\
0 & 2 & 0
\end{pmatrix}\, , \qquad {\bf k}_a = \begin{pmatrix}
0\\2\\0
\end{pmatrix}\, ,
\ee
and hence the scalings \eqref{ex3_w=2} agree with \eqref{genw2}.


\section{Conclusions}
\label{s:conclu}

In this work we have revisited the Emergence Proposal in Calabi--Yau compactifications of type II string theory, with particular emphasis on the role of light axionic strings, or EFT strings, in them. We have focused on the vector multiplet sector of type IIA large-volume compactifications, where such EFT strings are made of NS5-branes wrapping Nef divisors of the Calabi--Yau $X_6$. These objects define a representative set of infinite distance limits in vector multiplet moduli space. For each limit, the corresponding EFT string becomes asymptotically tensionless, together with a tower of KK modes and one or several towers of BPS particles made up from D-branes wrapping internal cycles of $X_6$. The relation between the EFT string tension and the scale $m_*$ of the lightest tower is controlled by the so-called scaling weight $w$ as in \eqref{scaling}. This quantity can take the values $w=1,2,3$, it coincides with the singularity index of \cite{Grimm:2018ohb} and it serves to organise the different kinds of limits that we have analysed. 

We have first focused on understanding the main differences between $w=3$ and $w=2$ limits. In the first case there is essentially a single leading tower of BPS particles becoming asymptotically massless, that of D0-branes. In the second case there is instead a double tower, because D2-branes wrapping a particular curve of $X_6$ become massless as fast as the D0's, and so do the bound states of both. This turns out to be a crucial fact, because while for testing the SDC it suffices to focus on a single leading tower, in order to check the Emergence Proposal it is important to have control over the full spectrum of particles that lie below the species scale. Indeed, taking this double tower into account prompts the species scale to decrease faster, and eventually reproduces the asymptotic behaviour of all the axion-independent components of the gauge kinetic function, correcting the apparent mismatch found in \cite{Grimm:2018ohb}. 

There are several interesting points regarding this result. First, it is remarkable that all $w=3$ limits have a single leading tower of light particles, while $w=2$ limits always feature a double tower.\footnote{Up to the possibility for $w=2$ limits outlined in footnote \ref{caveat1}, for which we do not have a geometric interpretation.} Following \cite{Lee:2019wij}, one can see that any $w=3$ limit correspond to a decompactification towards M-theory on $X_6$, with no further decompactification. One can then interpret the double tower of $w=2$ limits as a decompactification towards F-theory on $X_6$ with no further decompactification, which would be in agreement with the results of \cite{Corvilain:2018lgw,Lee:2019wij}. This would  explain the fact that, in both cases, the EFT string scale $\Lambda_{\rm NS5} = \sqrt{\cal T}$ coincides with the species scale. Indeed, in both pictures the NS5-brane becomes a 5d or 6d string with finite tension, and so it sets a scale proportional to the fundamental scale of M/F-theory. 

It is also worth mentioning one of the main lessons that we have drawn from the analysis of $w=2$ cases. For the Emergence Proposal to work in our setup, the leading multi-tower of states must be made up of charged particles. In other words, almost all species below $\Lambda_{\rm sp}$ must be charged. On hindsight, this is to be expected  for 4d $\CN=2$ theories like the ones that we are analysing, since they feature a direct relation between the field space metric and the gauge kinetic functions. Thus, if one wants to describe the asymptotic behaviour of both quantities via one-loop threshold corrections, only charged particles can respect the $\CN=2$ relations. In fact, in $w=3$ and $w=2$ limits, such charged particles are BPS, which are the obvious candidates to generate correlated corrections for the gauge kinetic function and the field space metric. 

We have then turned our attention to $w=1$ limits, which are conceptually different from the rest, because they feature several kinds of towers with leading asymptotic scaling, see table \ref{tab:scaling_L(phi)}. Among these appears the tower of EFT string oscillations, and so a priori one should consider two possible very different scenarios. If on the one hand the EFT string corresponds to a non-critical string, then one expects a finite number of oscillation modes. Then the spectrum of light states will be dominated by BPS charged particles made of D4/D2/D0 bound states. This is a more involved setup than in $w=2$ limits, but conceptually alike, and it reproduces the expected asymptotic behaviour of the gauge kinetic functions in a similar way. If on the other hand the EFT string is critical, as it is the general expectation,  the light spectrum of particles will be dominated by its oscillation modes. As argued in section \ref{s:eslim} some of these modes must be charged, and in fact they display the same charges as BPS particles made from D-branes. This is a crucial observation for the Emergence Proposal, because it allows for a string spectrum that is dominated by charged modes, as one would like to require based on the lesson learnt in $w=2$ limits. We have then tested the Emergence Proposal in this scenario, using a standard Ansatz for the degeneracy of modes at each level, and imposing the hypothesis of charged-state domination. One then finds that the Emergence Proposal is confirmed also in this case, at least for the gauge kinetic couplings. It would be interesting to extend our analysis to the corrections to the field space metric, to see if they follow under the same hypothesis, or if they require some additional assumption regarding the distribution of light BPS states. Of course, it would also be very important to verify our assumptions by considering explicit spectra of dual emergent critical strings. If they were not correct, it could either mean that the Emergence Proposal is seriously challenged, or that it is realised in a completely different way. 

Independently of the criticality of the string, one interesting feature of $w=1$ limits is that one not only can reproduce divergences of the gauge kinetic function via the Emergence Proposal, but also those components that asymptotically tend to zero. These zeroes are interpreted as divergences in the kinetic terms of the magnetic dual frame, sourced by light states with the charge of D4-branes. While for $\CN=2$ theories this behaviour could have been guessed, the analogue in setups with lower supersymmetry should correspond to divergences of the inverse of the moduli space metric. If the Emergence Proposal could predict these and other  divergences of the metric in general, it could lead to a more precise characterisation of the field space metric near the boundaries of the moduli space.

Finally, it would be interesting to see how the results of this paper could be extended. An obvious extension would be to the hyper multiplet sector of type II Calabi--Yau compactifications, where multi-towers were observed for instantons \cite{Marchesano:2019ifh,Grimm:2019wtx,Baume:2019sry} and to other $\CN=2$ compactifications beyond type II on Calabi--Yau manifolds. Another interesting direction would be to connect with the results of \cite{Gopakumar:1998ii,Gopakumar:1998jq} and to consider compactifications to different  dimensions, along the lines of \cite{Castellano:2022bvr}, and with lower supersymmetry. From our results and those in \cite{Lee:2019wij} it would seem that the Emergence Proposal is a vehicle to connect an infinite distance limit to a dual frame. If this was found to be a general statement, it could mean that the network of infinite distance limits is in fact a manifestation of the duality web.

\bigskip

\bigskip

\centerline{\bf  Acknowledgments}

\vspace*{.5cm}

We thank Alberto Castellano,  Damian van de Heisteeg, \'Alvaro Herr\'aez, Luis E. Ib\'a\~nez, Luca Martucci, Eran, Palti, Irene Valenzuela, Timo Weigand and Max Wiesner for discussions.  This work is supported through the grants CEX2020-001007-S and PID2021-123017NB-I00, funded by MCIN/AEI/10.13039/501100011033 and by ERDF A way of making Europe. LM is supported by the fellowship LCF/BQ/DI21/11860035 from "la Caixa" Foundation (ID 100010434).


\appendix


\section{Democratic formulation of  type IIA  on a Calabi--Yau}
\label{ap:democratic}

In this section we review the democratic formulation of 10-dimensional type IIA supergravity \cite{Bergshoeff:2001pv} and we show how to recover the 4-dimensional effective action \eqref{SVM} upon compactification on a Calabi--Yau. The democratic formulation in 10 dimensions is described by the pseudo-action
\be \label{10d_paction}
\begin{split}
S_{\rm IIA}^{(10)} &= \frac{1}{2\kappa_{10}^2} \int_{\IR^{1,3} \times X_6} e^{-2\phi} (R^{10} \ast_{10} 1 + 4d\phi \wedge \ast_{10} d\phi) \\
&- \frac{1}{8\kappa_{10}^2} \int_{\IR^{1,3} \times X_6} \Big( 2e^{-2\phi} H_3 \wedge \ast_{10} H_3 + \sum_{p=1}^4 G_{2p} \wedge *_{10} G_{2p} \Big)\, ,
\end{split}
\ee
where
\be
G_{2} = dC_1 \, , \qquad G_{2p}= dC_{2p-1} - H_{3} \wedge C_{2p-3}\, , \qquad H_3 = dB_2\, .
\ee
Since this is a pseudo-action, in order to recover the dynamics of type IIA supergravity, we have to impose on the equations of motion the following duality relations
\be \label{10d_duality}
G_6 = - \ast_{10} G_4\, , \qquad G_8 = \ast_{10} G_2\, .
\ee
The pseudo-action is invariant (up to total derivatives) under the following 10d gauge transformations
\bes \label{10d_gauge_tfm}
\begin{align}
&\begin{cases}
C_{2p-1} \rightarrow C_{2p-1} + d\Lambda_{2p-2}\\
C_{2p+1} \rightarrow C_{2p+1} + B_2 \wedge d\Lambda_{2p-2}
\end{cases}\, ,\\
&C_7 \rightarrow C_7 + d\Lambda_6\, ,\\
&B_2 \rightarrow B_2 + d\Lambda_1\, .
\end{align}
\ees
We aim to reduce this pseudo-action over a Calabi--Yau manifold $X_6$. We are particularly interested in the vector bosons that arise in 4d from reducing the different RR potentials, so we will focus on the last term of the pseudo-action. Let us then make the following decompositions:
\be
 C_{1} = \hat{A}^{0}  \, , \qquad C_3 = \hat{A}^{a} \wedge \omega_a\, ,  \qquad C_5 = \hat{A}_a \wedge \tilde{\omega}^a\, ,  \qquad C_7 = \hat{A}_0 \wedge \omega_6\, ,
\ee
where all the $\hat{A}$'s are Minkowski one-forms, $\omega_a$, $\tilde{\omega}^a$ is a basis of harmonic two-forms and four-forms such that $\int_{X_6} \omega_a \wedge \tilde{\omega}^b = \delta_a^b$, and finally $\int_{X_6} \omega_6 = 1$. The B-field axions are defined as
\be
B = b^a \omega_a\, ,
\ee
where $b^a$ are Minkowski scalars. As we will see in the following, it is useful to define the new $U(1)$ gauge fields
\bes
\label{basisu1}
\begin{align}
A^0 & = \hat{A}^0\, ,\\
A^a & = \hat{A}^a - b^a  \hat{A}^0\, , \\
A_a & = \hat{A}_a - \CK_{abc} \hat{A}^b b^c + \oh \CK_{abc} b^b b^c \hat{A}^0\, ,\\
A_0 & = -\hat{A}_0 + b^a \hat{A}_a - \oh \CK_{abc} b^b b^c \hat{A}^a + \frac{1}{6} \CK_{abc} b^a b^b b^c  \hat{A}^0\, ,
\end{align}
\ees
where $\CK_{abc} = \ell_s^{-6}\int_{X_6} \omega_a \wedge \omega_b \wedge \omega_c$.\footnote{In our conventions, the CY volume form is given by $\frac{1}{6} J \wedge J \wedge J$.} Let us denote the 4d field strengths of these gauge bosons by $F^A = dA^A$ and $F_A = dA_A$, where $A=(0,a)$. Then we have that
\bes
\label{GFs}
\begin{align}
G_2 & = d\hat{A}^0 =  F^0 \, ,\\
G_4 & = \left(d\hat{A}^a - db^a \wedge \hat{A}^0\right) \wedge \omega_a = \left(F^a + b^a F^0 \right) \wedge \omega_a  \, , \\
G_6 & = \left(d \hat{A}_a - \CK_{abc} db^b \wedge \hat{A}^c \right) \wedge \tilde{\omega}^a= \left( F_a + \CK_{abc} b^b F^c + \oh \CK_{abc}b^bb^c F^0 \right)  \wedge \tilde{\omega}^a \, ,\\
G_8 & = \left(d\hat{A}_0 - db^a \wedge \hat{A}_a \right) \wedge \omega_6  = \left( -F_0 + b^a F_a + \oh \CK_{abc} b^a b^b F^c  + \frac{1}{6} \CK_{abc} b^a b^b b^c  F^0 \right) \wedge \omega_6\, .
\end{align}
\ees
Plugging these expressions into \eqref{10d_paction} and integrating over $X_6$, we get the 4d pseudo-action, which can be written as
\be
S_{\rm IIA} \supset  -\frac{1}{8\kappa_4^2} \int_{M_4}  {\bf F}^t {\cR}^t {\cal G} {\cR} \ast_4{\bf F}\, ,
\ee
where we defined a new vector of field strengths ${\bf F}^t = (F^0, F^a, F_a ,-F_0)$ and
\be \label{CGCR}
\CG \, =\, \frac{\CK}{6}
\left(
\begin{array}{cccc}
 1  \\ 
  & 4 g_{ab} \\ 
  && \frac{9}{\CK^2} g^{ab} \\
  & & &   \frac{36}{\CK^2}
\end{array}
\right)\, , \qquad 
\cR = 
\left(
\begin{array}{cccc}
 1 &  \\ 
  b^i & \delta^{i}_{j} \\ 
 \oh  \CK_{ijk}b^jb^k &  \CK_{ijk}b^k &   \delta^{i}_{j} & 0 \\
\frac{1}{6} \CK_{ijk}b^ib^jb^k  & \oh  \CK_{ijk}b^jb^k& b^i & 1
\end{array}
\right)\, .
\ee
Alternatively, one can rewrite this expression as
\be \label{4d_paction}
S_{\rm IIA} \supset \frac{1}{8\kappa_{4}^2} \int_{M_4}  {\bf \bar{F}}^t
\begin{pmatrix}
\mathbb{I} & 0\\
-R & \mathbb{I}
\end{pmatrix}
\begin{pmatrix}
I^{-1} & 0\\
0 & I
\end{pmatrix}
\begin{pmatrix}
\mathbb{I} & -R\\
0 & \mathbb{I}
\end{pmatrix}
\ast_4 {\bf \bar{F}}\, ,
\ee
where we arranged the newly defined 4d field strengths in a vector ${\bf \bar{F}}^t = (F_0, F_a, F^0, F^a)$, and the matrices $I$ and $R$ are those that appear in the original action \eqref{SVM}
\be \label{I,R}
I = -\frac{\CK}{6}
\begin{pmatrix}
1+4g_{ab}b^a b^b & 4g_{ab} b^a\\
4g_{ab} b^b & 4g_{ab}
\end{pmatrix},
\quad R = -
\begin{pmatrix}
\frac{1}{3} \CK_{abc} b^a b^b b^c & \frac{1}{2} \CK_{abc} b^a b^c\\
\frac{1}{2} \CK_{abc} b^b b^c & \CK_{abc} b^c
\end{pmatrix}\, .
\ee

Substituting \eqref{GFs} into \eqref{10d_duality} we can express the duality relations in terms of the 4d fields
\bes    \label{dual_fs}
\begin{align}
F_a &= -\frac{2\CK}{3} g_{ab} \ast_4 (F^b + b^b F^0) - \CK_{abc} b^b \left(F^c +\frac{1}{2} b^c F^0\right)\, ,\\
F_0 &= -\frac{\CK}{6} \ast_4 F^0 + b^a F_a +\frac{1}{2} \CK_{abc} b^a b^b \left(F^c + \frac{1}{3} b^c F^0\right) \notag\\
&= -\frac{\CK}{6} (1+4g_{ab}b^a b^b) \ast_4 F^0 -\frac{2\CK}{3} g_{ab} b^a \ast_4 F^b -\frac{1}{2} \CK_{abc} b^a b^b F^c - \frac{1}{3} \CK_{abc} b^a b^b b^c F^0\, .
\end{align}
\ees
Computing the equations of motion from \eqref{4d_paction} and using the duality relations we get
\bes    \label{4d_EOM}
\begin{align}
&d\Big[\frac{\CK}{6} (1+4g_{ab}b^a b^b) \ast_4 F^0 + \frac{2\CK}{3} g_{ab} b^a \ast_4 F^b + \frac{1}{2} \CK_{abc} b^a b^b F^c + \frac{1}{3} \CK_{abc} b^a b^b b^c F^0\Big] = -dF_0 = 0\, ,\\
&d\Big[\frac{2\CK}{3} g_{ab}b^b \ast_4 F^0 + \frac{2\CK}{3} g_{ab} \ast_4 F^b + \CK_{abc} b^b F^c + \frac{1}{2} \CK_{abc} b^b b^c F^0\Big] = -dF_a = 0\, ,
\end{align}
\ees
and
\bes    \label{4d_Bianchi}
\begin{align}
dF^a &= 0\, ,\\
dF^0 &= 0\, .
\end{align}
\ees
These are to be interpreted, respectively, as the equations of motion and Bianchi identities for the 4d gauge fields, once we have expressed the dual fields in terms of the fundamental ones. In particular \eqref{4d_EOM} are the equations of motion we get from the following action
\be   \label{4d_IIA_action}
S_{\rm IIA}^{(4)} = \frac{1}{2\kappa_4^2} \int \Big[ R^{(4)} \ast_4 1 - 2g_{ab} \, dT^a \wedge \ast_4 d\bar{T}^b + I_{AB} \, F^A \wedge \ast_4 F^B + R_{AB} \, F^A \wedge F^B \Big]\, ,
\ee
where $I$ and $R$ are defined in \eqref{I,R}, which is the standard form of 4d $\cN=1$ supergravity action, see e.g. \cite{Sabra:2015tsa}. In addition, the dual field strengths $F_A=(F_0,F_a)$ in \eqref{dual_fs} coincide with the duals defined from this 4d action as
\be
G_A \equiv \frac{\delta S}{\delta F^A} = I_{AB} \ast_4 F^B + R_{AB} F^B\, .
\ee
This action is gauge invariant (up to total derivatives) under the 4d version of the gauge transformations \eqref{10d_gauge_tfm}, namely
\bes   \label{4d_gauge_tfm}
\begin{align}
&A^0 \raw A^0 + d\lambda_0\, ,\\
&\begin{cases}
b^a \rightarrow b^a + n^a\\
A^a \rightarrow A^a - n^a A^0
\end{cases},\\
&A^a \raw A^a + d\lambda^a\, .
\end{align}
\ees
Note, in particular, that the field strengths $F^a$ are not gauge invariant, while  $F^a + b^a F^0$ are.

\section{Curvature-corrected kinetic terms}
\label{ap:K3}

In this appendix we write the complete expressions of the type IIA action \eqref{SVM} and pseudo-action \eqref{psSVM} containing the curvature corrections encoded in the prepotential \eqref{corr_prep}
\begin{equation}
{\cal F} = -\frac{1}{6} {\cal K}_{abc}\frac{Z^a Z^b Z^c}{Z^0} + \frac{1}{2} K_{ab}^{(1)}Z^a Z^b + K_{a}^{(2)} Z^a Z^0 + \frac{i}{2} K^{(3)} (Z^0)^2\, , 
\end{equation}
where we switched to homogeneous coordinates $Z^A=(Z^0,Z^a)$, such that $T^a \equiv \frac{Z^a}{Z^0}$.
From the prepotential one can compute the curvature-corrected gauge kinetic matrix $\bar{I}_{AB} \equiv \text{Im} \, {\cal N}_{AB}$ and the matrix in the topological term $\bar{R}_{AB} \equiv \text{Re} \, {\cal N}_{AB}$, using
\be
{\cal N}_{AB} = \bar{{\cal F}}_{AB} + 2i \frac{\text{Im}{\cal F}_{AC} Z^C \text{Im}{\cal F}_{BD} Z^D}{Z^C \text{Im}{\cal F}_{CD} Z^D}\, ,
\ee
where ${\cal F}_{AB} = \frac{\p^2 {\cal F}}{\p Z^A \p Z^B}$. In order to match with our conventions, we need to redefine the graviphoton field strength by flipping its sign $F^0 \rightarrow -F^0$, so the off-diagonal terms ${\cal N}_{0A}$ and ${\cal N}_{A0}$ computed using the previous equation pick up a minus sign. After doing this the curvature-corrected matrices $\bar{I}$ and $\bar{R}$ read
\be
\bar{I} = -
\left(
\begin{array}{cc}
\frac{\tilde{\cal K}}{6} \frac{1+3\vee}{1+\frac{3}{4}\vee} + \left[ \frac{2}{3} \tilde{\cal K} \tilde{g}_{ab} -\frac{3}{2 \tilde{\cal K}} {\cal K}_a {\cal K}_b \frac{9\vee}{4+3\vee} \right] b^a b^b  & \left[ \frac{2}{3} \tilde{\cal K} \tilde{g}_{ab} -\frac{3}{2 \tilde{\cal K}} {\cal K}_a {\cal K}_b \frac{9\vee}{4+3\vee} \right] b^b \\ 
\left[ \frac{2}{3} \tilde{\cal K} \tilde{g}_{ab} -\frac{3}{2 \tilde{\cal K}} {\cal K}_a {\cal K}_b \frac{9\vee}{4+3\vee} \right] b^b & \frac{2}{3} \tilde{\cal K} \tilde{g}_{ab} -\frac{3}{2 \tilde{\cal K}} {\cal K}_a {\cal K}_b \frac{9\vee}{4+3\vee}
\end{array}
\right)\, ,
\ee
\be
\bar{R} = - 
\left(
\begin{array}{cc}
 \frac{1}{3} {\cal K}_{abc} b^a b^b b^c  + \frac{9\vee}{4+3\vee} {\cal K}_a b^a & \frac{1}{2} {\cal K}_{abc}b^b b^c + \frac{1}{2} \frac{9\vee}{4+3\vee} {\cal K}_a + K_a^{(2)} \\ 
\frac{1}{2} {\cal K}_{abc}b^b b^c + \frac{1}{2} \frac{9\vee}{4+3\vee} {\cal K}_a + K_a^{(2)} & {\cal K}_{abc}b^c - K_{ab}^{(1)}
\end{array}
\right)\, ,
\ee
where we defined $\vee \equiv \frac{K^{(3)}}{\cal K}$ and introduced the modified volume and moduli space metric
\be
\tilde{V}_{X} = \frac{\tilde{\cal K}}{6} = \frac{\cal K}{6} \left( 1-\frac{3}{2} \vee \right)\, ,
\ee
\be
\tilde{g}_{ab} = \frac{3}{2\tilde{\cal K}} \left( \frac{3}{2\tilde{\cal K}} {\cal K}_a {\cal K}_b - {\cal K}_{ab} \right)\, .
\ee
These expressions enter in the action \eqref{SVM}, while if we rotate to the $U(1)$ basis $\tilde{F}^A$, as in \eqref{SVMg}, the matrices read
\be
\tilde{I} =
\left(
\begin{array}{cc}
\frac{\tilde{\cal K}}{6} \frac{1+3\vee}{1+\frac{3}{4}\vee} & 0 \\ 
0 & \frac{2}{3} \tilde{\cal K} \tilde{g}_{ab} -\frac{3}{2 \tilde{\cal K}} {\cal K}_a {\cal K}_b \frac{9\vee}{4+3\vee}
\end{array}
\right)\, ,
\ee
\be
\tilde{R} =
\left(
\begin{array}{cc}
 \frac{1}{3} {\cal K}_{abc} b^a b^b b^c -2 K_a^{(2)}b^a - K_{ab}^{(1)}b^a b^b & -\frac{1}{2} {\cal K}_{abc}b^b b^c + K_a^{(2)} + K_{ab}^{(1)}b^b + \frac{1}{2} \frac{9\vee}{4+3\vee} {\cal K}_a \\ 
-\frac{1}{2} {\cal K}_{abc}b^b b^c + K_a^{(2)} + K_{ab}^{(1)}b^b + \frac{1}{2} \frac{9\vee}{4+3\vee} {\cal K}_a & {\cal K}_{abc}b^c - K_{ab}^{(1)}
\end{array}
\right)\, .
\ee
Let us now turn to the pseudo-action \eqref{psSVM}, which contains the following corrected matrices
\be
\bar{\cal R} = {\cal R} Q =
\left(
\begin{array}{cccc}
1 & 0 & 0 & 0 \\ 
b^a & \delta^{a}_{b} & 0 & 0 \\ 
\frac{1}{2} {\cal K}_{abc}b^b b^c + K_a^{(2)} &  {\cal K}_{abc}b^c - K_{ab}^{(1)} &   \delta^{a}_{b} & 0 \\
\frac{1}{6} {\cal K}_{abc} b^a b^b b^c + b^a K_a^{(2)} & \frac{1}{2} {\cal K}_{abc} b^b b^c - K_{ab}^{(1)} b^b - K_a^{(2)} & b^a & 1
\end{array}
\right)\, ,
\ee
\be
\bar{\cal{G}} =
\left(
\begin{array}{cccc}
\frac{\tilde{\cal K}}{6} \left( 1+\frac{9\vee(1+3\vee)}{(2-3\vee)^2} \right) & 0 & \frac{1}{2}t^a \frac{9\vee}{2-3\vee} & 0 \\ 
0 & \frac{2}{3}\tilde{\cal K}\tilde{g}_{ab} - \frac{27}{8\tilde{\cal K}} {\cal K}_a {\cal K}_b \frac{\vee}{1+3\vee} & 0 & -\frac{3}{2\tilde{\cal K}} {\cal K}_a \frac{9\vee}{2(1+3\vee)} \\
\frac{1}{2}t^a \frac{9\vee}{2-3\vee} & 0 & \frac{3}{2\tilde{\cal K}} \left( \tilde{g}^{ab} + t^at^b \frac{9\vee}{1+3\vee} \right)  & 0 \\
0 & -\frac{3}{2\tilde{\cal K}} {\cal K}_a \frac{9\vee}{2(1+3\vee)} & 0 & \frac{6}{\tilde{\cal K}} \left( 1-\frac{9\vee}{4(1+3\vee)} \right)
\end{array}
\right)\, .
\ee
Notice that the monodromy matrix $\bar{\cal{R}}$ only depends on the axions $b^a$ and on the curvature corrections $K_{ab}^{(1)}$ and $K_a^{(2)}$, while the saxions $t^a$ and the correction $K^{(3)}$ only enter $\bar{\cal G}$, which now assumes a non-diagonal form. Notice also that, as expected, if we send $\vee \rightarrow 0$ all these expressions reduce to the ones in Section \ref{s:scaling}, which were computed neglecting $K^{(3)}$.



\bibliographystyle{JHEP2015}
\bibliography{papers}

\end{document}